\documentclass[reqno,a4paper,11pt]{article}
\pdfoutput=1
\usepackage{xcolor}

\usepackage{graphicx}
\usepackage[textwidth = 430 pt, textheight = 630 pt]{geometry}

\definecolor{MyDarkBlue}{rgb}{0.15,0.25,0.45}
\usepackage{epsfig,rotating}
\usepackage{amsmath,amssymb}
\usepackage{amsfonts}
\usepackage{mathrsfs}
\usepackage{bbm}
\usepackage[normalem]{ulem}
\usepackage[T1]{fontenc}

\usepackage{latexsym}
\usepackage{amsthm}
\usepackage[all,knot]{xy}
\xyoption{arc}

\usepackage[utf8x]{inputenc}

\usepackage{hyperref}
\hypersetup{
hypertexnames=false,
colorlinks=true,
citecolor=MyDarkBlue,
linkcolor=MyDarkBlue,
urlcolor=MyDarkBlue,
pdfauthor={},
pdftitle={},
pdfsubject={hep-th math-ph}
breaklinks=true
}

\usepackage{tikz}
\usepackage{mathtools}

\let\fn\footnote
\renewcommand{\footnote}[1]{\linespread{1.1}\fn{#1}\linespread{1.29}}

\makeatletter\renewcommand{\section}{\@startsection
{section}{1}{\z@}{-3.5ex plus -1ex minus
    -.2ex}{2.3ex plus .2ex}{\bf }}
\makeatletter\renewcommand{\subsection}{\@startsection{subsection}{2}{\z@}{-3.25ex
plus -1ex minus
   -.2ex}{1.5ex plus .2ex}{\bf }}
\makeatletter\renewcommand{\subsubsection}{\@startsection{subsubsection}{3}{-2.45ex}{-3.25ex
plus -1ex minus -.2ex}{1.5ex plus .2ex}{\it }}
\renewcommand{\thesection}{\arabic{section}}
\renewcommand{\thesubsection}{\arabic{section}.\arabic{subsection}}
\renewcommand{\@seccntformat}[1]{\@nameuse{the#1}.~~}

\renewcommand{\theequation}{\thesection.\arabic{equation}}
\makeatletter \@addtoreset{equation}{section}

\providecommand*{\xmapstofill@}{%
  \arrowfill@{\mapstochar\relbar}\relbar\rightarrow
}
\providecommand*{\xmapsto}[2][]{%
  \ext@arrow 0395\xmapstofill@{#1}{#2}%
}

\providecommand*{\xhookrightfill@}{%
  \arrowfill@{\lhook\joinrel\relbar}\relbar\rightarrow
}
\providecommand*{\xhookrightarrow}[2][]{%
  \ext@arrow 0395\xhookrightfill@{#1}{#2}%
}

\usepackage[toc,page]{appendix}

\newtheorem{thm}{Theorem}[section]
\renewcommand{\thethm}{\thesection.\arabic{thm}}

\newtheorem{remark}[thm]{Remark}

\newcommand{\myxymatrix}[1]{\vcenter{\vbox{\xymatrix{#1}}}}

\renewcommand{\appendices}{
\section*{Appendix}\label{appendices}\setcounter{subsection}{0}
\addcontentsline{toc}{section}{Appendix}
\setcounter{equation}{0}
\makeatletter
\renewcommand{\theequation}{\Alph{subsection}.\arabic{equation}}
\renewcommand{\thesubsection}{\Alph{subsection}}
\renewcommand{\thethm}{\Alph{subsection}.\arabic{thm}}
\@addtoreset{equation}{subsection}
\@addtoreset{thm}{subsection}
\makeatother
}

\def\slasha#1{\setbox0=\hbox{$#1$}#1\hskip-\wd0\hbox to\wd0{\hss\sl/\/\hss}}

\def\periodb#1{\setbox0=\hbox{$#1$}#1\hskip-\wd0\hbox to\wd0{-}}

\newcommand{\unit}{\mathbbm{1}}   			
\newcommand{\im}{\mathrm{im}}   			
\newcommand{\id}{\mathrm{id}}   			

\newcommand{\CC}{\mathcal{C}}
\newcommand{\CCC}{\mathscr{C}}

\newcommand{\CF}{\mathcal{F}}

\newcommand{\CCG}{\mathscr{G}}
\newcommand{\CH}{\mathcal{H}}

\newcommand{\CI}{\mathcal{I}}

\newcommand{\CL}{\mathcal{L}}

\newcommand{\CQ}{\mathcal{Q}}

\newcommand{\CR}{\mathcal{R}}

\newcommand{\CT}{\mathcal{T}}

\newcommand{\CCX}{\mathscr{X}}

\newcommand{\CE}{\mathcal{E}}
\newcommand{\CCE}{\mathscr{E}}
\newcommand{\frg}{\mathfrak{g}}				

\newcommand{\FR}{\mathbbm{R}}     			
\newcommand{\NN}{\mathbbm{N}}     			
\newcommand{\RZ}{\mathbbm{Z}}     			

\newcommand{\dd}{\mathrm{d}}     			
\newcommand{\dpar}{\partial}     			
\newcommand{\embd}{{\hookrightarrow}}     		

\newcommand{\de}{\mathrm{e}}     			

\newcommand{\eand}{{\qquad\mbox{and}\qquad}}     		
\newcommand{\ewith}{{\qquad\mbox{with}\qquad}}

\newcommand{\der}[1]{\frac{\dpar}{\dpar #1}}   		
\newcommand{\derr}[2]{\frac{\dpar #1}{\dpar #2}}   	
\newcommand{\pr}{\mathsf{pr}}     			

\newcommand{\ao}{\mathfrak{o}}

\newcommand{\sU}{\mathsf{U}}     			

\newcommand{\sH}{\mathsf{H}}
\newcommand{\sL}{\mathsf{L}}
\newcommand{\sT}{\mathsf{T}}

\newcommand{\sLie}{\mathsf{Lie}}

\newcommand{\sO}{\mathsf{O}}

\newcommand{\sDiff}{\mathsf{Diff}}

\renewcommand{\remark}[1]{}     				
\def\tyng(#1){\hbox{\tiny$\yng(#1)$}}			
\def\tyoung(#1){\hbox{\tiny$\young(#1)$}}			

\newcommand{\beq}{\begin{eqnarray}}
\newcommand{\eeq}{\end{eqnarray}}

\begin{document}
\begin{titlepage}
\begin{flushright}
 EMPG--18--24
\end{flushright}
\vskip 1.0cm
\begin{center}
{\LARGE \bf Extended Riemannian Geometry III:\\[0.2cm] Global Double Field Theory with Nilmanifolds}
\vskip 1.cm
{\Large Andreas Deser$^{a}$ and Christian S\"amann$^b$}
\setcounter{footnote}{0}
\renewcommand{\thefootnote}{\arabic{thefootnote}}
\vskip 1cm
{\em${}^a$ Mathematical Institute of the Charles University \\
Sokolovsk{\' a} 83\\
186 75 Praha 8, Czech Republic
}\\[0.5cm]
{\em${}^b$
Department of Mathematics,
Heriot--Watt University\\
Colin Maclaurin Building, Riccarton, Edinburgh EH14 4AS, U.K.}\\
and\\  {\em Maxwell Institute for Mathematical Sciences, Edinburgh,
  U.K.} \\ and \\ {\em Higgs Centre for Theoretical Physics,
  Edinburgh, U.K.}\\[0.5cm]
{Email: {\ttfamily andreas3deser@gmail.com , c.saemann@hw.ac.uk}}
\end{center}
\vskip 1.0cm
\begin{center}
{\bf Abstract}
\end{center}
\begin{quote}
We describe the global geometry, symmetries and tensors for Double Field Theory over pairs of nilmanifolds with fluxes or gerbes. This is achieved by a rather straightforward application of a formalism we developed previously. This formalism constructs the analogue of a Courant algebroid over the correspondence space of a T-duality, using the language of graded manifolds, derived brackets and we use the description of nilmanifolds in terms of periodicity conditions rather than local patches. The strong section condition arises purely algebraically, and we show that for a particularly symmetric solution of this condition, we recover the Courant algebroids of both nilmanifolds with fluxes. We also discuss the finite, global symmetries of general local Double Field Theory and explain how this specializes to the case of T-duality between nilmanifolds.
\end{quote}
\end{titlepage}

\tableofcontents

\section{Introduction and results}

The goal of Double Field Theory is to provide a Lagrangian which is manifestly symmetric under the hidden $\sO(d,d;\RZ)$ T-duality symmetry of string theory and governs the dynamics of the massless sector of superstring theory, see~\cite{Aldazabal:2013sca,Berman:2013eva,Hohm:2013bwa} and references therein. Currently, the available Lagrangian and its gauge structure are given only on a local patch of target space\footnote{with minor exceptions; see e.g.~\cite{Cederwall:2014kxa,Cederwall:2014opa}, \cite{Blumenhagen:2014gva, Blumenhagen:2015zma,Hassler:2016srl} or also~\cite{Freidel:2017yuv} for alternative approaches to ours} and it is suggested that some form of gluing has to be performed to recover the global picture. Since T-duality is fundamentally linked to compact target space directions, a clean understanding of this gluing procedure is important.

In this paper, we study this problem in the case of the prominent example of 3-dimensional nilmanifolds. These are circle bundles over the torus $T^2$ and we equip them with a three form flux, which mathematically corresponds to the curvature of a gerbe. The advantage of this class of examples is that the non-trivial patchings can be understood in terms of twisted periodicity conditions on the local data. We will use the general formalism we presented previously in~\cite{Deser:2016qkw} and extend the local picture developed there to a global one by demanding compatibility with the periodicity conditions. 

In physics terms, we are studying the NS-NS sector of supergravity without dilaton\footnote{A generalization to heterotic NS-NS gravity along the lines discussed in~\cite{Deser:2017fko} should be straightforward.}, consisting of a vielbein (or metric) on a topological manifold $M$ carrying an abelian gerbe with connective structure (i.e.~Kalb--Ramond field) $B$ and curvature (i.e.~flux) $H=\dd B$. The symmetries of this theory consist of diffeomorphisms and gauge transformations of the $B$-field together with the gauge-for-gauge transformations of the latter. At the infinitesimal level, these are elegantly described by Hitchin's {\em Generalized Geometry}, cf.~e.g.~\cite{Gualtieri:2003dx}. The vector field and the 1-form parameterizing these symmetries are sections of the {\em generalized tangent bundle}, which has some non-trivial gluing properties if the gerbe is topologically non-trivial. 

The perspective of the generalized tangent bundle can be further improved by considering the corresponding Courant algebroid in the language of symplectic differential graded manifolds or {\em $Q$-manifolds} as pioneered in~\cite{Roytenberg:0203110}. The infinitesimal symmetry structure of NS-NS gravity is then simply obtained in terms of derived brackets, cf.~\cite{Deser:2014mxa, Deser:2018oyg}. From a more mathematical perspective, abelian gerbes are categorified principal bundles or principal 2-bundles for short. For each topological class of abelian gerbes, there is a corresponding class of Courant algebroids (and Courant algebroids are best regarded as symplectic Lie 2-algebroids). The Lie 2-algebra\footnote{By this term, we shall always mean a 2-term $L_\infty$-algebra, which is essentially the same.} of symmetries of a gerbe is then the associated Lie 2-algebra of the corresponding Courant algebroid, obtained from the derived bracket construction of~\cite{Fiorenza:0601312,Getzler:1010.5859}. 

The aim is now to lift this rather well-understood global picture from Generalized Geometry to Double Field Theory. The latter associates additional winding degrees of freedom to the compact circle fibers of the nilmanifold. Locally, T-duality is understood as an exchange of the coordinates Fourier-dual to the momentum and winding degrees of freedom but globally, we have topological features complicating the picture.

One approach to a global picture of Double Field Theory is to use the finite symmetry actions to glue together local descriptions~\cite{Berman:2014jba}. Recall that in the case of Generalized Geometry, the infinitesimal symmetries are governed by the Dorfman bracket. While the symmetries themselves form a Lie 2-algebra which is awkward to integrate, the {\em actions} of the infinitesimal symmetries by the Dorfman bracket form a Lie algebra, which integrates to the semidirect product of diffeomorphisms and closed 2-forms on the base manifold. In the case of Double Field Theory, the infinitesimal symmetries are governed by an analogue of the Dorfman bracket, often called the generalized Lie derivative or D-bracket. Again, the symmetries form a Lie 2-algebra, and the actions by the D-bracket form an ordinary Lie algebra which can be integrated to finite transformations~\cite{Hohm:2012gk}, see also~\cite{Hull:2014mxa} for further discussions and~\cite{Park:2013mpa} for a slightly different perspective. These finite transformations were then used as gluing transformations for local descriptions of Double Field Theory~\cite{Berman:2014jba}. 

There is, however, a problem with this construction, which is already seen in the context of Generalized Geometry. As stated above, each topological class of gerbes comes with its own class of Courant algebroids. Switching from a topologically trivial $H$-flux (i.e.~$H=\dd B$ globally) to a non-trivial one therefore requires either to modify the Dorfman bracket or to modify the gluing prescriptions for the corresponding generalized tangent bundle. Both changes are related by a coordinate change, as we shall see. Since the generalized Lie derivative in Double Field Theory is almost exclusively discussed without taking this modification into account\footnote{We are not aware of any discussion of this point before~\cite{Deser:2018oyg}.}, it is not surprising that the above mentioned integrated infinitesimal symmetries, as well as the gluing transformations, are only suitable for topologically trivial gerbes. The same conclusion was reached by explicit computations in~\cite{Papadopoulos:2014mxa}. 

In our previous paper~\cite{Deser:2018oyg}, we presented a formalism in which the necessary modifications of the D-bracket due to topologically non-trivial gerbes become natural and can be discussed in detail, see also~\cite{Heller:2016abk}. The key observation was that the $Q$-manifold picture of the Courant algebroid over space-time can be extended to an analogous picture over the target space of Double Field Theory. We showed that this requires to drop the condition $Q^2=0$, leading us to {\em symplectic pre-N$Q$-manifolds}. On these, we recover the generalized Lie derivative of Double Field Theory as a derived bracket, and the strong section condition (or rather a slight weakening thereof) is recovered algebraically by demanding that the derived brackets form a Lie 2-algebra of symmetries, as on the Courant algebroid. 

The modification of the Dorfman bracket of Generalized Geometry for topologically non-trivial gerbes with curvature $H$ is obtained by adding $H$ to the Hamiltonian $\CQ$ of the differential given by the homological vector field $Q=\{\CQ,-\}$. This modification has a clear analogue for the symplectic pre-N$Q$-manifolds underlying the structure of Double Field Theory.

The above was already claimed in~\cite{Deser:2018oyg}. In this paper, we show that this sketch can indeed be completed to a full coherent picture. We study the simple example of T-duality between 3-dimensional nilmanifolds carrying topologically non-trivial gerbes. This example is still simple enough to allow for a transparent and detailed discussion in coordinates. In particular, we can use the fact that nilmanifolds are the orbit spaces of particular $\RZ_3$-actions on $\FR^3$. Therefore, we do not have to use cumbersome local patches, but we can work with flat space and a number of periodicity conditions to describe functions and global sections of the various bundles. Note also that for more general examples, one needs to leave the realm of smooth manifolds and generalize to noncommutative and nonassociative spaces, which we plan to do in future work. 

Recall that three dimensional nilmanifolds $N_j$ are principal circle bundles over the 2-torus $T^2$, and thus they are characterized by an integer $j\in \RZ$, the Chern class of the bundle. Similarly, gerbes over $N_j$ are also given by an integer, their Dixmier--Douady class $[H]=k$, which is an element in $H^3(N_j,\RZ)\cong \RZ$.  Under T-duality along the fiber direction of the principal bundle, one obtains the nilmanifold $N_k$ endowed with a gerbe of Dixmier--Douady class $j$:
\begin{equation}
 \CCG_{j,k}=(N_j,[H]=k)~~\xleftrightarrow{~~\mbox{T-duality}~~}~~\CCG_{k,j}=(N_k,[H]=j)~,
\end{equation}
which follows from the Buscher rules and which was explained in detail in~\cite{Bouwknegt:2003vb}.

A classical way to understand T-duality is to use correspondence spaces, which are fibered products of original and T-dual manifold over a common base manifold. \emph{Locally}, we associate the original- and dual fiber coordinates to momentum and winding coordinates. In the case at hand, the correspondence space $K$ between $N_j$ and $N_k$ is the fiber product $K=N_j\times_{T^2} N_k$. The gerbes $\CCG_{j,k}$  and $\CCG_{k,j}$ come with associated Courant algebroids $\CCC_{j,k}$ and $\CCC_{k,j}$ (by which we always mean the corresponding $Q$-manifolds as given in~\cite{Roytenberg:0203110}), which encode the relevant Lie 2-algebra of symmetries.\footnote{A picture of T-duality as an isomorphism of invariant Courant algebroids was given in~\cite{Cavalcanti:2011wu}.} Both gerbes and both associated Courant algebroids can be pulled back to the correspondence space $K$. Tensoring the gerbes, we obtain the gerbe $\CCG_K$ with associated Courant algebroid $\CCC_K$. The space relevant for Double Field Theory is now a graded submanifold $\CCE_K$ of $\CCC_K$. Altogether, we have the following diagram:
\vspace*{0.8cm}
\begin{equation}
 \myxymatrix{
     & \CCG_K\ar@{->}[dr] \ar@{<.>}@/^4ex/[rr]& \CCE_K\ar@{->}[d] & \CCC_K \ar@{->}[dl] \ar@{->}[l] & & \\
     \CCC_{j,k}\ar@{<.>}[d] \ar@{->}[dr] & & K:=N_j\times_{T^2}N_k \ar@{->}[dl]^{\pr_1} \ar@{->}[dr]_{\pr_2}& & \CCC_{k,j} \ar@{->}[dl]\ar@{<.>}[d]\\
     \CCG_{j,k} \ar@{->}[r] & N_j\ar@{->}[dr]_{\pi_j} & & N_k\ar@{->}[dl]^{\pi_k} & \CCG_{k,j} \ar@{->}[l] \\
     & & T^2 & &
    }
\end{equation}
The symplectic pre-N$Q$-manifold $\CCE_K$ plays the same role in Double Field Theory as the Courant algebroid for Generalized Geometry. Its sections parameterize the symmetries whose structures and actions are governed by derived brackets. Our first important result is therefore the detailed description of $\CCE_K$ in terms of coordinates. Explicitly, we lift the periodicity conditions of the coordinates on the Courant algebroids $\CCC_{j,k}$ and $\CCC_{k,j}$ to periodicity conditions of the various coordinates on $\CCE_K$ and we show that the ingredients to construct the C-bracket via a derived bracket are invariant under this periodicity and therefore they are indeed global.

Having found $\CCE_K$, we can follow our formalism of~\cite{Deser:2018oyg} further and derive the conditions for the antisymmetrized derived brackets to form a Lie 2-algebra of symmetries. This Lie 2-algebra of symmetries will always appear in the form of a 2-term $L_\infty$-algebra.\footnote{Note that this 2-term $L_\infty$-algebra gives rise to, but is {\em not} identical to the $L_\infty$-algebra of Double Field Theory as partially given in~\cite{Hohm:2017pnh} which arises from the symmetry Lie 2-algebra via the classical BV formalism, as explained in~\cite{Jurco:2018sby}.} The resulting conditions are the appropriate global form of (a slight weakening of) the strong section condition for Double Field Theory making the T-duality between nilmanifolds carrying gerbes manifest.

Solutions to this section conditions are found in a very natural way. Given a Poisson structure with compatible vector field $Q$ and corresponding Hamiltonian $\CQ$, we have $Q^2F=\tfrac12\{\{\CQ,\CQ\},F\}$ for functions $F$ on $\CCE_K$ (actually on any pre-N$Q$-manifold of even degree). The expression $\{\CQ,\CQ\}$ now factorizes as
\begin{equation}
 \{\CQ,\CQ\}=p_3p_4~,
\end{equation}
where $p_3$ and $p_4$ are the momentum coordinates in the T-duality directions. They form Hamiltonians of the vector fields $\der{x^3}=\{p_3,-\}$ and $\der{x^4}=\{p_4,-\}$ along the lifts of the projections $\pr_1$ and $\pr_2$ to the Courant algebroids,
\begin{equation}
 \hat \pr_1:\CCE_K \rightarrow \CCC_{j,k}\eand\hat \pr_2:\CCE_K\rightarrow\CCC_{k,j}~.
\end{equation}
Complementing these by global vector fields $V_3$ and $V_4$, we arrive at two integrable distributions, $\langle \der{x^4}, V_4\rangle$ and $\langle \der{x^3},V_3\rangle$. The first defines $\hat \pr_1$ and the second one defines $\hat \pr_2$. Both projections admit particularly simple sections, which are embeddings $e_1:\CCC_{j,k}\embd \CCE_K$ and $e_2:\CCC_{k,j}\embd \CCE_K$ given by morphisms of pre-N$Q$-manifolds. That is, they are smooth maps of graded manifolds and the Hamiltonian of $Q$ as well as the symplectic form are pullbacks of the data on $\CCE_K$ along $e_1$ and $e_2$.

We note that some aspects of the specialization of our example in which $k\in \RZ$ but $j=0$ were discussed earlier in~\cite{Hull:2009sg} and the overlapping results agree.

Altogether, we conclude that there is no geometric obstacle to a global description of Double Field Theory, at least for those cases in which the manifest T-duality connects smooth manifolds carrying gerbes.

This paper is structured as follows. In section~2, we review nilmanifolds, gerbes over these and the corresponding Courant algebroids. The symmetry structures (both for infinitesimal and finite symmetries) of these Courant algebroids are then explained in detail in terms of derived brackets in section~3. In section~4, we present the Double Field Theory lift of this picture. We start with a very brief review of topological T-duality for nilmanifolds, as found in~\cite{Bouwknegt:2003vb} before we construct the symplectic pre-N$Q$-manifold $\CCE_K$ which encodes the geometry and symmetries of Double Field Theory. We discuss the strong section condition as well as the most relevant solutions, restricting $\CCE_K$ to the Courant algebroids of the two nilmanifolds with gerbes. We also make some detailed comments on extended differential geometry. In~\cite{Deser:2018oyg}, we had introduced the concepts of extended tensors, extended covariant derivatives, extended torsion and Riemann tensors. All this is readily specialized to $\CCE_K$, and the extended tensors of $\CCE_K$ reduce nicely to those of the two Courant algebroids $\CCC_{j,k}$ and $\CCC_{k,j}$ for the corresponding solutions of the strong section condition. Section~5 contains a discussion of the infinitesimal and finite symmetries of general Double Field Theory before we specialize to the case discussed in section~4. We close in section~6 with a brief discussion of generalizations of our constructions.

\section{Geometry of gerbes on three-dimensional nilmanifolds}

\subsection{Three-dimensional nilmanifolds}

A {\em nilmanifold} $M$ is a compact quotient manifold $M=N/\Gamma$, where $N$ is a nilpotent Lie group and $\Gamma$ a discrete subgroup in $N$. Recall that a {\em nilpotent Lie group} is connected and has Lie algebra $\frg$ such that the sequence
\begin{equation}
 \frg_1=[\frg,\frg]~,~~~\frg_2=[\frg_1,\frg]~,~~~\dots
\end{equation}
becomes trivial at some term $\frg_k$.

We shall be interested in examples of 3-dimensional nilmanifolds\footnote{These examples play a prominent role in Thurston's geometrization conjecture: every closed 3-manifold can be decomposed into pieces with one of eight geometric structures, one being that of a nilmanifold.}. Recall that up to isomorphisms, there is a unique, connected, simply connected nilpotent Lie group, which is the Heisenberg group $\sH(3)$ of upper-triangular $3\times 3$-matrices. The cocompact discrete subgroups $\Gamma_j$ are classified by an integer $j\in\NN$. We shall denote the nilmanifold $\sH(3)/\Gamma_j$ by $N_j$.

The nilmanifold $N_j$ is a circle bundle over $T^2$, and we shall describe it in terms of coordinates $x^i$ on $\FR^3$ with the following periodicity conditions imposed:
\begin{equation}\label{eq:coords_nilmanifold}
 (x^1,x^2,x^3)\sim (x^1,x^2+1,x^3)\sim (x^1,x^2,x^3+1)\sim (x^1+1,x^2,x^3-jx^2)~.
\end{equation}
Here, $x^1$ and $x^2$ are coordinates on $T^2$, while $x^3$ is the coordinate along the fiber. Clearly, we can include the case $j=0$, for which the circle bundle becomes trivial: $N_0=T^2\times S^1=T^3$.

We have a basis of globally defined 1-forms\footnote{The bar is merely distinguishing these tensors from a slightly different set $(\xi^i,\zeta_i)$ used mostly in the later discussion.}
\begin{subequations}\label{eq:global_vec_forms_Nj}
\begin{equation}
 \bar \xi^1=\dd x^1~,~~~\bar \xi^2=\dd x^2~,~~~\bar \xi^3=\dd x^3+jx^1\dd x^2
\end{equation}
together with dual vector fields 
\begin{equation}
 \bar \zeta_1=\der{x^1}~,~~~\bar \zeta_2=\der{x^2}-jx^1\der{x^3}~,~~~\bar \zeta_3=\der{x^3}
\end{equation}
\end{subequations}
satisfying $\langle \bar \xi^i,\bar \zeta_j\rangle= \delta^i_j$.

The circle bundle $\pi_j:N_j\rightarrow T^2$ can be described using a single transition function $h$ specifying the transition $x^1\rightarrow x^1+1$. In its additive form, it reads as $h=-jx^2$, and we can extend this \v Cech cocycle to a Deligne cocycle $(h,A)$, encoding a principal circle bundle with connection $A$. One compatible choice is the connection $A=jx^1\dd x^2$, which satisfies the cocycle condition $A(x^1+1)-A(x^1)=-\dd h$. The resulting curvature reads as $F=j\dd x^1\wedge \dd x^2$ and thus $j$ is the first Chern number $j=\int_{T^2} F$ of the principal circle bundle $N_j$.

Finally, we note that the nilmanifold $N_j$ comes with the natural metric
\begin{equation}\label{eq:metric_nilmanifold}
 g=(\dd x^1)^2+(\dd x^2)^2+(\dd x^3+jx^1\dd x^2)^2~.
\end{equation}

\subsection{Gerbes on three-dimensional nilmanifolds}

At the heart of Generalized Geometry and T-duality is the fact that the metric $g$ is complemented by a 2-form $B$ which is part of the connective structure of a gerbe. For our purposes, the rather simplistic definition of gerbes in the form of {\em Hitchin--Chatterjee gerbes}~\cite{Hitchin:1999fh} will be sufficient. 

Consider a manifold $M$ together with a cover $\sqcup_a U_a\twoheadrightarrow M$. Let $U_{ab}:=U_a\cap U_b$, $U_{abc}:=U_a\cap U_b\cap U_c$, etc. Then a {\em gerbe}, or {\em $\sU(1)$-gerbe}, or {\em Hitchin--Chatterjee gerbe}, is a \v Cech 2-cocycle $h$, that is a collection of functions 
\begin{equation}\label{eq:cocycle1}
 h_{(abc)}:U_a\cap U_b\cap U_c \rightarrow \FR\ewith h_{(abc)}-h_{(abd)}+h_{(acd)}-h_{(bcd)}=0~~~\mbox{on}~~~U_{abcd}~.
\end{equation}
Just as in the case of principal circle bundles, we can extend the \v Cech cocycle encoding the gerbe to a Deligne cocycle encoding a gerbe with {\em connective structure}, the analogue of a connection for a gerbe. Such a Deligne cocycle consists of the \v Cech cocycle $h_{(abc)}$ on triple overlaps $U_{abc}$, 1-forms $A_{(ab)}$ on the double overlaps $U_{ab}$ and 2-forms $B_{(a)}$ on the patches $U_a$. The cocycle conditions~\eqref{eq:cocycle1} are then completed by
\begin{equation}\label{eq:cocycles_gerbe}
 A_{(ab)}-A_{(ac)}+A_{(bc)}=-\dd h_{(abc)}~~\mbox{on}~~U_{abc}\eand B_{(a)}-B_{(b)}=\dd A_{(ab)}~~\mbox{on}~~U_{ab}~.
\end{equation}
The second cocycle condition implies that the curvatures on each $U_a$, 
\begin{equation}
 H|_{U_a}=\dd B_{(a)}~,
\end{equation}
form indeed a global 3-form. Just as the curvature 2-form $F$ of a principal $\sU(1)$-bundle captures the first Chern class\footnote{i.e.~up to torsion elements} $c_1(P)=[F]$, so the curvature 3-from $H$ of a gerbe $\CCG$ encodes its Dixmier--Douady class $dd(\CCG)=[H]$.

Geometrically, gerbes should be regarded as central groupoid extensions. That is, they are particular categories in which both the morphisms and the objects form manifolds. A rather detailed review of $\sU(1)$-gerbes and the relation between the various definitions is found, e.g., in~\cite{Bunk:2016rta}.

Just as principal $\sU(1)$-bundles, gerbes can be tensored and restricted to submanifolds. Geometrically, this amounts to tensoring and restricting their categorified fibers. From the cocycle perspective, tensoring amounts to adding the underlying cocycles, yielding a gerbe whose Dixmier--Douady class is the sum of the Dixmier--Douady classes of the individual gerbes.\footnote{More explicitly, the two gerbes on $M$ are described by cocycles with respect to two covers of $M$. There is a common refinement of these two covers, over which we can add the corresponding cocycles.} A restriction to some submanifold is also readily performed by restricting the cocycles to that submanifold.\footnote{Explicitly, given a submanifold $X\embd M$ of some manifold $M$, the cocycles on $M$ are given with respect to some cover $U_M$. Together with the embedding $X\embd M$, this cover induces a cover $U_X$ on $X$ and we can restrict the cocycles on $M$ by pullback.}

Let us now come to gerbes on the nilmanifolds $N_j$. The Dixmier--Douady classes of these are encoded in global closed 3-forms, and since $N_j$ is connected, compact, orientable and 3-dimensional, these forms are unique up to rescaling. The cocycle conditions impose a quantization condition on the Dixmier--Douady class, just as it is the case for the Chern class for principal $\sU(1)$-bundles. Altogether, gerbes on $N_j$ are characterized by an integer $k$, and we denote the corresponding gerbe by $\CCG_{j,k}$.

The 3-form representing the Dixmier--Douady class of $\CCG_{j,k}$ is the global 3-form
\begin{equation}
 H\ =\ k\,\dd x^1\wedge \dd x^2\wedge \dd x^3~,~~~\int_{N_j} H=k~,
\end{equation}
where we used again coordinates~\eqref{eq:coords_nilmanifold}. Note that a 2-form $B$ giving rise to this curvature is
\begin{equation}\label{eq:B_field_nilmanifold}
 B\ =\ k\,x^1 \dd x^2\wedge \dd x^3~.
\end{equation}
Clearly, this 2-form is not global because $B(x^1+1)-B(x^1)=k\,\dd x^2\wedge \dd x^3\neq 0$. The cocycle conditions~\eqref{eq:cocycles_gerbe} then implies that
\begin{equation}\label{eq:dA}
 \dd A=k\,\dd x^2\wedge \dd x^3~.
\end{equation}

\subsection{Exact Courant algebroids and generalized metrics}

An {\em exact Courant algebroid} over a manifold $M$ is usually defined as a vector bundle $E$ which fits into the short exact sequence of vector bundles,
\begin{equation}\label{eq:Courant_sequence}
 0 \xrightarrow{~\phantom{\rho}~} T^*M \xhookrightarrow{~\phantom{\rho}~} E \xrightarrow{~\rho~} TM \xrightarrow{~\phantom{\rho}~} 0~,
\end{equation}
where $\rho:E\rightarrow TM$ is called the {\em anchor map}, and which is endowed with a fiberwise inner product $\langle -,-\rangle:E\times E\rightarrow M\times \FR$ and a bracket $\mu_2(-,-):\Gamma E\times \Gamma E\rightarrow \Gamma E$ called {\em Courant bracket}, subject to a number of axioms. 

A {\em splitting} of the anchor map $\rho$, i.e.~a bundle map $\sigma:TM\embd E$ such that $\rho\circ \sigma=\id$, yields the projector $P:=\sigma\circ \rho$ and thus an isomorphism
\begin{equation}
 E=\im(P)\oplus \ker(P)\cong TM\oplus T^*M~.
\end{equation}
Explicitly, consider the anchor map
\begin{equation}
 \rho:E\cong TM\oplus T^*M\rightarrow TM~,~~~\rho(X+\alpha)=X~.
\end{equation}
A general splitting is then of the form
\begin{equation}
 \sigma~: X~\mapsto~X+\omega (X)=X+(g+B)(X)~,
\end{equation}
where we split the rank~2 tensor $\omega$ into its symmetric and antisymmetric parts $g$ and $B$. It also leads to the two graphs
\begin{equation}\label{eq:graphs}
 C_\pm=\{X+(\pm g+B)(X)\,|\,X\in \Gamma(TM)\}~,
\end{equation}
which span the subbundles of $E$ on which the inner product
\begin{equation}
 \langle X+\alpha,Y+\beta\rangle=\alpha(Y)+\beta(X)
\end{equation}
is positive and negative definite, respectively. Such a half-dimensional subbundle $C_+$ of $E$ with positive definite induced metric is called a {\em generalized metric}. More explicitly, let $G$ be an endomorphism on $E$ such that 
\begin{equation}\label{eq:ker_condition}
 C_\pm=\ker(G\mp\unit)\eand G^2=GG^T=\unit
\end{equation}
and let $\eta$ be the $\sO(d,d)$-metric
\begin{equation}
\eta=\begin{pmatrix} 0 & \unit \\ \unit & 0 \end{pmatrix}\ewith
 \langle X+\alpha, Y+\beta\rangle~=~
 \begin{pmatrix}X \\ \alpha\end{pmatrix}^T
 \eta
 \begin{pmatrix}Y \\ \beta\end{pmatrix}~.
\end{equation}
Then the map $\CH:\Gamma(E)\rightarrow \Gamma(E^*)$ with $\CH=\eta G$ is also called the {\em generalized metric}. With $C_\pm$ as given in~\eqref{eq:graphs} and~\eqref{eq:ker_condition}, it follows that
\begin{equation}
 G=\left(\begin{array}{cc}
 -g^{-1}B & g^{-1} \\
 g-Bg^{-1}B & B g^{-1}          
         \end{array}\right)\eand
 \CH=\left(\begin{array}{cc}
 g-Bg^{-1}B & B g^{-1} \\
 -g^{-1}B & g^{-1} 
         \end{array}\right)~.
\end{equation}

\subsection{Exact Courant algebroids as symplectic Lie 2-algebroids}

A more powerful perspective is obtained by regarding a Courant algebroid as a symplectic Lie 2-algebroid $\CCC$ or {\em symplectic N$Q$-manifold of degree~2} as explained in detail in~\cite{Roytenberg:0203110}. That is, $\CCC$ is an $\NN$-graded manifold with body $M$, which carries a symplectic form of $\NN$-degree~2 as well as a nilquadratic vector field $Q$ of degree~1 generating a symplectomorphism. For a very detailed discussion of symplectic Lie $n$-algebroids in our conventions, see~\cite{Deser:2016qkw}. 

As shown in~\cite{Roytenberg:0203110}, the N$Q$-manifold underlying an exact Courant algebroid over a manifold $M$ is isomorphic to $\CCC=T^*[2]T[1]M$. This N$Q$-manifold is essentially the vector bundle $T^*TM$, but regarded as a graded manifold with local coordinates $x^\mu$, $\xi^\mu$, $\zeta_\mu$, $p_\mu$ of degrees~0, 1, 1 and 2, respectively. Locally, the functions on $\CCC$ are polynomials in the fiber coordinates with coefficients which are arbitrary functions of the $x^\mu$. There is a canonical symplectic form and a canonical vector field of degree~1 in the above coordinates,
\begin{equation}
 \omega=\dd x^\mu\wedge \dd p_\mu+\dd \xi^\mu\wedge \dd \zeta_\mu~,~~~Q=\xi^\mu \der{x^\mu}+p_\mu\der{\zeta_\mu}
\end{equation}
and $Q$ satisfies indeed $Q^2=0$ and $\CL_Q\omega=0$. The symplectic form $\omega$ induces a Poisson structure\footnote{See again~\cite{Deser:2016qkw} for details and our conventions.},
\begin{equation}\label{eq:Poisson_bracket}
\begin{aligned}
 \{f,g\}&:=\left(\der{p_\mu} f\right)\left(\der{x^\mu} g\right)-\left(\der{x^\mu}f\right)\left(\der{p_\mu} g\right)\\
 &\hspace{1cm}-(-1)^{|f|}\left(\der{\zeta_\mu}f\right)\left(\der{\xi^\mu}g\right)-(-1)^{|f|}\left(\der{\xi^\mu}f\right)\left(\der{\zeta_\mu} g\right)~,
\end{aligned}
\end{equation}
and $Q$ is Hamiltonian:
\begin{equation}\label{eq:canonical_Q}
 Q=\{\CQ,-\}\ewith \CQ=\xi^\mu p_\mu~.
\end{equation}

The relation between the classical definition of an exact Courant algebroid and the N$Q$-manifold picture is now as follows. Putting $p_\mu=0$ on $\CCC$ yields a vector bundle over $M$ with fibers of degree~1, $E[1]\rightarrow M$. Here, $E[1]$ is the grade-shifted bundle $E$ in~\eqref{eq:Courant_sequence}. Functions $\CC^\infty(\CCC)$ of degree~1 correspond to sections of $E$ and in terms of our local coordinates, we identify
\begin{equation}
 X+\alpha=X^\mu(x)\zeta_\mu+\alpha_\mu(x)\xi^\mu~~~\longleftrightarrow~~~X^\mu(x)\der{x^\mu}+\alpha_\mu(x)\dd x^\mu\in \Gamma(TM\oplus T^*M)~.
\end{equation}
The kernel of the anchor map are the sections spanned by the coefficient of $p_\mu$ in $\CQ$, which is here $\xi^\mu$. Explicitly, the anchor map reads as $\rho(X^\mu\zeta_\mu)=X^\mu\der{x^\mu}$. The pairing on $E$ is given by the Poisson bracket~\eqref{eq:Poisson_bracket} on functions of degree~1 on $\CCC$ and the Courant bracket,
\begin{equation}
 \mu_2(X_1+\alpha_1,X_2+\alpha_2)=\tfrac12\left(\{\{\CQ,X_1+\alpha_1\},X_2+\alpha_2\}-\{\{\CQ,X_2+\alpha_2\},X_1+\alpha_1\}\right)~,
\end{equation}
is part of an associated Lie 2-algebra that canonically arises on $\CCC$ from the homological vector field $\CQ$. We shall discuss this Lie 2-algebra in more detail in section~\ref{sec:global_symmetries}.

\subsection{The Courant algebroid of a gerbe}

Consider now a gerbe $\CCG$ with connective structure $(A_{(ab)}, B_{(a)})$ with respect to a cover $\sqcup_a U_a\twoheadrightarrow M$ of $M$ and let $H=\dd B\in H^3(M)$ be a representative of its Dixmier--Douady class $dd(\CCG)$. This data induces local splits $\sigma$ of the anchor map $\rho:E[1]\rightarrow TM$, given over the patch $U_a$ by
\begin{equation}\label{eq:splitting}
 \sigma\left(\der{x^\mu}\right):=\der{x^\mu}+B_{\mu\nu}\dd x^\nu~,~~~B_{(a)}=\tfrac12 B_{\mu\nu}\dd x^\mu\wedge \dd x^\nu~.
\end{equation}
If we further insert a symmetric tensor, $g_{\mu\nu}$,
\begin{equation}\label{eq:splitting2}
 \sigma\left(\der{x^\mu}\right):=\der{x^\mu}+(g_{\mu\nu}+B_{\mu\nu})\dd x^\nu~,
\end{equation}
we can capture any splitting, as discussed above. For now, however, we restrict ourselves to splittings of the form~\eqref{eq:splitting}. 

In the symplectic N$Q$-picture, equation~\eqref{eq:splitting} translates to maps
\begin{equation}
 \sigma: TU_a \rightarrow \CCC^\infty_1(T^*[2]T[1]U_a)~,~~~\sigma\left(\der{x^\mu_{(a)}}\right):=\zeta^{(a)}_\mu+B^{(a)}_{\mu\nu}\xi^\nu_{(a)}~.
\end{equation}
On the intersection of patches, we then have the relation
\begin{equation}\label{eq:glue_zeta}
 \zeta_\mu^{(a)}=\zeta_\mu^{(b)}+\dpar_{[\mu} A^{(ab)}_{\nu]}\xi^\nu~,
\end{equation}
because of the Deligne cocycle relation $B_{(a)}-B_{(b)}=\dd A_{(ab)}$, cf.~\eqref{eq:cocycles_gerbe}. In terms of sections $X+\alpha$ of $TM\oplus T^*M$, this relation corresponds to the well-known patching relation for the {\em twisted generalized tangent bundle},
\begin{equation}\label{eq:patching-k}
 X_{(a)}+\alpha_{(a)} = X_{(b)}+\alpha_{(b)}+\iota_{X_{(b)}}\dd A_{ab}~.
\end{equation}

The coordinate change $\zeta_\mu\rightarrow \zeta_\mu+B_{\mu\nu}\xi^\nu$ induced by the splitting is still compatible with the original symplectic form and homological vector field. (The necessary relations, $Q^2=0$ and $\CL_Q\omega=0$ are local conditions, and $Q$ and $\omega$ are now simply glued together in a different way.) One can, however, perform a coordinate change in which the $B$-field contribution to the coordinate transformations is removed but instead one has a change in the homological vector field $Q$, which is best expressed by a modification of its Hamiltonian $\CQ$,
\begin{equation}\label{eq:twist}
 \CQ=\xi^\mu p_\mu+\tfrac{1}{3!}H_{\mu\nu\kappa}\xi^\mu\xi^\nu\xi^\kappa~.
\end{equation}
Here, $H=\dd B=\tfrac{1}{3!} H_{\mu\nu\kappa}\dd x^\mu\wedge \dd x^\nu\wedge \dd x^\kappa$. We shall see an explicit example of this transformation later. In both cases, we arrive at the {\em twisted Courant algebroid} $\CCC_\CCG$.

Note that in~\eqref{eq:twist}, the condition for $\{\CQ,\CQ\}=0$, which is equivalent to $Q^2=0$, is $\dd H=0$. Moreover, one can show that the additional contributions from the $B$-field can be transformed away by a symplectomorphism on $\CCC$ if and only if $H$ is exact, see e.g.~\cite{Deser:2016qkw}. Similarly, splittings~\eqref{eq:splitting2} can be brought to the form~\eqref{eq:splitting} by a symplectomorphism. Thus, relevant for the twist is indeed the cohomology class of $[H]$, the Dixmier--Douady class $dd(\CCG)$ of the gerbe $\CCG$. We conclude that the isomorphism classes of Courant algebroids over $M$ are determined by $H^3(M)$ and the corresponding characteristic class is called the {\em \v Severa class} of the Courant algebroid. 

We thus obtain a one-to-one correspondence between Courant algebroids $\CCC_\CCG$ and gerbes $\CCG$, up to isomorphism classes, mapping the Dixmier--Douady class to the \v Severa class. As discussed in~\cite{Deser:2016qkw}, the Courant algebroid $\CCC_\CCG$ comes with an associated Lie 2-algebra which describes the symmetries of the gerbe $\CCG$. In this sense, the Courant algebroid generalizes the Atiyah algebroid of principal $\sU(1)$-bundles, cf.~\cite{Bressler:2002ur,Collier:1108.1525}.

In the twisted case, a {\em generalized metric} is still a half-dimensional subbundle $C_+$ of $E$ with positive definite induced metric. Over patches $U_a$, we still have
\begin{equation}
 C_\pm|_{U_a}=\{X+(\pm g+B)(X) | X\in \Gamma(TU_a)\}~,
\end{equation}
but $E$ and its subbundles $C_\pm$ will have non-trivial transition functions, in general. This leads to the endomorphism $G$ and the generalized metric to be defined only locally:
\begin{equation}
 G_{(a)}=\left(\begin{array}{cc}
 -g^{-1}B_{(a)} & g^{-1} \\
 g-B_{(a)}g^{-1}B_{(a)} & B_{(a)} g^{-1}          
         \end{array}\right)~,~~~
 \CH_{(a)}=\left(\begin{array}{cc}
 g-B_{(a)}g^{-1}B_{(a)} & B_{(a)} g^{-1} \\
 -g^{-1}B_{(a)} & g^{-1} 
         \end{array}\right)~.
\end{equation}
Decomposing $\CH_{(a)}$ as 
\begin{equation}\label{eq:B-field_trafo}
 \CH_{(a)}=\left(\begin{array}{cc} \unit & B_{(a)} \\ 0 & \unit \end{array}\right)\left(\begin{array}{cc} g & 0 \\ 0 & g^{-1} \end{array}\right)\left(\begin{array}{cc} \unit & B_{(a)} \\ 0 & \unit \end{array}\right)^T~,
\end{equation}
we readily verify that 
\begin{equation}\label{eq:gen_metric_transition}
 \CH_{(a)}=\left(\begin{array}{cc} \unit & \dd A_{(ab)} \\ 0 & \unit \end{array}\right)\CH_{(b)}\left(\begin{array}{cc} \unit & \dd A_{ab} \\ 0 & \unit \end{array}\right)^T
\end{equation}
due to $B_{(a)}=B_{(b)}+\dd A_{(ab)}$ and we obtain the transition function for the generalized metric.

\subsection{\mathversion{bold}The Courant algebroid of \texorpdfstring{$\CCG_{j,k}$}{G(j,k)}}

We now give the explicit Courant algebroid $\CCC_{j,k}:=\CCC_{\CCG_{j,k}}$ for the gerbe $\CCG_{j,k}$ over the nilmanifold $N_j$ with Dixmier--Douady class $k$. We use again the periodic coordinates $x^i$, $i=1,2,3$, with the identification
\begin{equation}
 (x^1,x^2,x^3)\sim (x^1,x^2+1,x^3)\sim (x^1,x^2,x^3+1)\sim (x^1+1,x^2,x^3-jx^2)~.
\end{equation}

Let us first consider the case $k=0$. We have much freedom in choosing the coordinates $\xi^i$ and $\zeta_i$ of degree~1 on $E[1]$ and it is sensible to consider Darboux coordinates, for which the symplectic form reads as
\begin{equation}
 \omega=\dd x^i\wedge \dd p_i+\dd \xi^i\wedge \dd \zeta_i~.
\end{equation}

By considering patches on $N_j$, we can use the exact Courant algebroid as described above and glue it together over patches. This leads to the coordinate system $(x^i, \tilde \xi^i, \tilde \zeta_i,\tilde p_i)$ with the anchor map and embedding 
\begin{equation}
\begin{gathered}
 \tilde \zeta_1\mapsto\der{x^1}~,~~~\tilde \zeta_2\mapsto\der{x^2}~,~~~\tilde \zeta_3\mapsto\der{x^3}~,\\
  \tilde \xi^1\leftrightarrow\dd x^1~,~~~\tilde \xi^2\leftrightarrow\dd x^2~,~~~\tilde \xi^3\leftrightarrow\dd x^3~.
\end{gathered}
\end{equation}
Using~\cite[Equation~3.3]{Roytenberg:0203110}, we then deduce that the identification of coordinates is
\begin{multline}
 (x^1\,,\,x^2\,,\,x^3\,,\,\tilde \xi^1\,,\,\tilde \xi^2\,,\,\tilde \xi^3\,,\,\tilde \zeta_1\,,\,\tilde \zeta_2\,,\,\tilde \zeta_3\,,\,\tilde p_1\,,\,\tilde p_2\,,\,\tilde p_3)\\
 =(x^1+1\,,\,x^2\,,\,x^3-jx^2\,,\,\tilde \xi^1\,,\,\tilde \xi^2\,,\,\tilde \xi^3-j\tilde \xi^2\,,\,\tilde \zeta_1\,,\, \tilde \zeta_2 + j\tilde \zeta_3\,,\, \tilde \zeta_3\,,\,\tilde p_1\,,\,\tilde p_2+j\tilde p_3\,,\,\tilde p_4)~.
\end{multline}
The symplectic form $\omega$ and the Hamiltonian $\CQ_0$ are the canonical ones,
\begin{equation}
 \omega=\dd x^i\wedge \dd \tilde p_i+\dd \tilde \xi^i\wedge \dd \tilde \zeta_i\eand \CQ_0 =\tilde \xi^i \tilde p_i~.
\end{equation}
As necessary, both expressions are indeed global.

Much more convenient, however, is the coordinate system $(x^i,\bar \xi^i, \bar \zeta_i, \bar p_i)$, where we use the global vector fields and dual 1-forms~\eqref{eq:global_vec_forms_Nj}. The anchor map thus reads as 
\begin{equation}
 \bar \zeta_1\mapsto\der{x^1}~,~~~\bar \zeta_2\mapsto\der{x^2}-jx^1\der{x^3}~,~~~\bar\zeta_3\mapsto\der{x^3}~~~\mbox{or}~~~ \bar \zeta_i\mapsto \tau^j_i \der{x^j}~,
\end{equation}
where we introduced the vielbein
\begin{equation}\label{eq:def_vielbein3d}
 \tau_i^j=\delta_i^j-jx^1\delta_i^2\delta^j_3~,
\end{equation}
not making any distinction between coordinate and non-coordinate indices. Since $\bar \xi^i$ and $\bar \zeta_i$ are supposed to be Darboux coordinates, the embedding $T^*M\embd E$ identifies
\begin{equation}
 \bar \xi^1\leftrightarrow\dd x^1~,~~~\bar \xi^2\leftrightarrow\dd x^2~,~~~\bar \xi^3\leftrightarrow\dd x^3+jx^1\dd x^2~~~\mbox{or}~~~ \bar \xi^j\leftrightarrow (\tau^{-1})^j_i \dd x^i~.
\end{equation}
Together with $\bar p_i= \tilde p_i$, we arrive at the following identification of coordinates:
\begin{multline}\label{coordperio}
 (x^1\,,\,x^2\,,\,x^3\,,\,\bar \xi^1\,,\,\bar \xi^2\,,\,\bar \xi^3\,,\,\bar \zeta_1\,,\,\bar \zeta_2\,,\,\bar \zeta_3\,,\,\bar p_1\,,\,\bar p_2\,,\,\bar p_3)\\
 =(x^1+1\,,\,x^2\,,\,x^3-jx^2\,,\,\bar \xi^1\,,\,\bar \xi^2\,,\,\bar \xi^3\,,\,\bar \zeta_1\,,\, \bar \zeta_2\,,\, \bar \zeta_3\,,\,\bar p_1\,,\,\bar p_2+j\bar p_3\,,\,\bar p_3)~.
\end{multline}
The appropriate global symplectic form $\omega$ and Hamiltonian $\CQ_0$ are then 
\begin{equation}
 \omega=\dd x^i\wedge \dd \bar p_i+\dd \bar \xi^i\wedge \dd \bar \zeta_i\eand \CQ_0=\bar\xi^i \bar p_i-j x^1\bar \xi^2 \bar p_3=\bar \xi^i\tau_i^j \bar p_j~,
\end{equation}

To include a non-trivial Dixmier--Douady class $dd(\CCG)=k$ of a gerbe $\CCG$, we have to modify the coordinates $\bar \xi^i$ and $\bar \zeta_i$, taking into account the gluing condition~\eqref{eq:patching-k}. From~\eqref{eq:dA} and~\eqref{eq:glue_zeta}, we glean that the transition from $x^1$ to $x^1+1$ should read as
\begin{equation}\label{Gerbeperio}
 X^i\bar \zeta_i+\alpha_i \bar \xi^i~~~\longrightarrow~~~X^i\bar \zeta_i+\alpha_i \bar \xi^i +k(X^2\bar \xi^3-X^3\bar \xi^2)~.
\end{equation}
Therefore, we have the identification
\begin{equation}
 (x^1\,,\,\bar \xi^1\,,\,\bar \xi^2\,,\,\bar \xi^3\,,\,\bar \zeta_1\,,\,\bar \zeta_2\,,\,\bar \zeta_3)~\sim~(x^1+1\,,\,\bar \xi^1\,,\,\bar \xi^2\,,\,\bar \xi^3\,,\,\bar \zeta_1\,,\,\bar \zeta_2+k\bar \xi^3\,,\,\bar \zeta_3-k\bar \xi^2)~. 
\end{equation}
We first observe that there is no additional periodicity condition for the momenta\linebreak from~\eqref{Gerbeperio}, but this is due to the fact that the transformation does not depend on the body coordinates. In the general case, also the momenta receive an additional periodicity condition due to the topological non-triviality of the gerbe. Both, $\omega$ and $\CQ_0$ remain invariant under the combined~\eqref{coordperio} and~\eqref{Gerbeperio} transformations. Hence, the expressions are globally defined and since $\{\CQ_0,\CQ_0\}=0$ as before, they complete the structure of a symplectic N$Q$-manifold and therefore the structure of a Courant algebroid. Note that we can compute a generalized metric from the natural metric on the nilmanifold~\eqref{eq:metric_nilmanifold} and the $B$-field for the gerbe $\CCG_{j,k}$,~\eqref{eq:B_field_nilmanifold}. The result is 
\begin{equation}
 \begin{pmatrix}
      1 & 0 & 0 & 0 & 0 & 0 \\
      0 & (1+j^2(x^1)^2)(1+k^2(x^1)^2) & jx^1(1+k^2(x^1)^2) & 0 & -jk(x^1)^2 & k x^1(1+j^2(x^1)^2)\\
      0 & j x^1(1+k^2(x^1)^2 & 1+k^2(x^1)^2 & 0 & -kx^1 & jk(x^1)^2\\
      0 & 0 & 0 & 1 & 0 & 0 \\
      0 & -jk(x^1)^2 & -kx^1 & 0 & 1 & -jx^1 \\
      0 & kx^1(1+j^2(x^1)^2) & jk (x^1)^2 & 0 & -jx^1 & 1+j^2(x^1)^2
     \end{pmatrix}~.
\end{equation}
In the transition $x^1\rightarrow x^1+1$, this generalized metric transforms indeed as~\eqref{eq:gen_metric_transition}.

Our coordinates of degree~1, $\bar \xi^i$ and $\bar \zeta_i$ are no longer global, as a non-trivial $B$-field induces the non-trivial gluing of the generalized tangent bundle~\eqref{eq:patching-k}. We certainly prefer, however, to work with coordinates $(x^i,\xi^i,\zeta_i,p_i)$, where the degree~1 members are global. It is well-known in Generalized Geometry, that the non-trivial gluing can be absorbed by performing a $B$-field transformation on the sections of $TM\oplus T^*M$, cf.~e.g.~\cite{Hitchin:2005in}. This amounts to
\begin{subequations}
\begin{equation}\label{eq:eB-coord_trafo_gtb}
 (\xi,\zeta)=\de^B(\bar \xi,\bar \zeta):= \de^{\lbrace -,B\rbrace }(\bar \xi,\bar \zeta)=\left(1-k x^1 \bar \xi^3\der{\bar \zeta_2}+k x^1 \bar \xi^2\der{\bar \zeta_3}\right)(\bar \xi,\bar \zeta)
\end{equation}
on the basis vectors. In terms of coordinates on the symplectic N$Q$-manifold, we have
\begin{equation}
x^i=\bar x^i~,~~~\xi^i=\bar \xi^i~,~~~\zeta_1=\bar \zeta_1~,~~~\zeta_2=\bar\zeta_2-kx^1\bar \xi^3~,~~~\zeta_3=\bar \zeta_3+kx^1\bar\xi^2~,
\end{equation}
which is complemented by 
\begin{equation}
 p_1=\bar p_1-k\bar \xi^2\bar \xi^3~,~~~p_2=\bar p_2~,~~~p_3=\bar p_3
\end{equation}
\end{subequations}
to a symplectomorphism, cf.~again~\cite[Equation~3.3]{Roytenberg:0203110}:
\begin{equation}
 \omega=\dd x^i\wedge \dd p_i+\dd \xi^i\wedge \dd \zeta_i~.
\end{equation}
For the new coordinate system, we have the identification
\begin{multline}
 (x^1\,,\,x^2\,,\,x^3\,,\,\xi^1\,,\,\xi^2\,,\,\xi^3\,,\,\zeta_1\,,\,\zeta_2\,,\,\zeta_3\,,\,p_1\,,\,p_2\,,\,p_3)\\
 =(x^1+1\,,\,x^2\,,\,x^3-jx^2\,,\,\xi^1\,,\,\xi^2\,,\,\xi^3\,,\,\zeta_1\,,\, \zeta_2\,,\, \zeta_3\,,\,p_1\,,\,p_2+jp_3\,,\,p_3)~.
\end{multline}
In the following, we shall use essentially exclusively the coordinates $(x^i,\xi^i,\zeta_i,p_i)$.

The Hamiltonian in the new coordinates reads as
\begin{equation}\label{eq:Courant_Hamiltonian}
 \CQ=\xi^i p_i-j x^1\xi^2 p_3+k\xi^1\xi^2\xi^3
\end{equation}
and we see that the dependence on the Dixmier--Douady class which was previously encoded in the transition function of the coordinates is now moved into the Hamiltonian $\CQ$. As one readily shows, there is no symplectomorphism eliminating $k$, and the \v Severa class $k$ describes the isomorphism classes of the symplectic N$Q$-manifold $\CCC_{j,k}$.

As a side remark, we observe that $\CQ$ is obtained from $\CQ_0$ by acting with $e^B:=\de^{\{-,B\}}$ with $B=\tfrac12 B_{\mu\nu}\xi^\mu\xi^\nu$: $\CQ = e^B\CQ_0$. This can be used to relate $\CQ_0$-closed functions which represent lifts of generalized vectors (i.e.~functions of degree 1) to $\CQ$-closed lifts of generalized vectors: if $X\in \CC^{\infty}_1(\CCC_{j,k})$ is $\CQ_0$-closed, then $e^BX$ is $\CQ$-closed,
\begin{equation}
\begin{aligned}
\lbrace \CQ, X\rbrace &= \lbrace e^B \CQ_0,e^B X\rbrace\\
&=\lbrace \CQ_0, X\rbrace + \Bigl\lbrace \lbrace \CQ_0,B\rbrace ,X\Bigr\rbrace + \Bigl\lbrace\CQ_0,\lbrace X,B\rbrace\Bigr\rbrace\\
&=\lbrace \CQ_0,X\rbrace + \Bigl\lbrace \lbrace \CQ_0,X\rbrace,B\Bigr\rbrace\\
&=\de^B\{\CQ_0,X\}~.
\end{aligned}
\end{equation}
This result is well-known in the classical language: it relates ordinary and twisted cohomology on the generalized tangent bundle~\cite{Hitchin:2005in}.

\section{Finite global symmetries of Generalized Geometry}\label{sec:global_symmetries}

One of the major advantages of the above formulation of Generalized Geometry is that it makes the relevant symmetry structures very obvious. As first observed in~\cite{Fiorenza:0601312,Getzler:1010.5859}, symplectic Lie $n$-algebroids always come with an associated Lie $n$-algebra structure, see also~\cite{Deser:2016qkw} for more details. The symmetry Lie 2-algebra of the gerbe $\CCG_{j,k}$ is given by the associated Lie 2-algebra of the Courant algebroid $\CCC_{j,k}$, cf.~\cite{Deser:2016qkw}.\footnote{This relation readily extends to $n$-gerbes.} The Courant algebroid is to the abelian gerbe roughly what the Atiyah algebroid is to a principal circle bundle, see e.g.~\cite{Bressler:2002ur,Collier:1108.1525}. We start with a description of the Lie 2-algebra associated to $\CCC_{j,k}$, before we interpret it as the appropriate symmetry Lie 2-algebra.

\subsection{Lie 2-algebra associated to \texorpdfstring{$\CCC_{j,k}$}{C(j,k)}}

The Lie 2-algebra associated to $\CCC_{j,k}$ has an underlying graded vector space $\sL=\sL_{-1}\oplus\sL_0$, where the homogeneous parts
\begin{equation}
 \sL_{-1}=\CC^\infty_0(\CCC_{j,k})=\CC^\infty(N_j)\eand \sL_0=\CC^\infty_1(\CCC_{j,k})\cong\Gamma(E)~,
\end{equation}
are the vector spaces of functions on $\CCC_{j,k}$ of degree~0 and 1, respectively. These correspond to ordinary functions on the base $N_j$ of the Courant algebroid as well as generalized vectors. The graded vector space $\sL$ becomes a Lie 2-algebra (a 2-term $L_\infty$-algebra) if we endow it with the following totally anti-symmetric brackets $\mu_i$ of degree~$2-i$:
\begin{equation}\label{eq:ass_Courant_algebra}
\begin{aligned}
 \mu_1(f)&= \{\CQ, f\}~,\\
 \mu_2(X+\alpha,Y+\beta)&=\tfrac12\big(\{\{\CQ,X^i\zeta_i+\alpha_i\xi^i\},Y^\nu\zeta_\nu+\beta_\nu\xi^\nu\}- (X+\alpha)\leftrightarrow(Y+\beta)\big)\\
 \mu_2(X+\alpha,f)&=\tfrac12\{\{\CQ,X+\alpha\},f\}~,\\
 \mu_3(X+\alpha,Y+\beta,Z+\gamma)&=\tfrac{1}{3!}\big(\{\{\{\CQ,X+\alpha\},Y+\beta\},Z+\gamma\}+\dots\big)\\
 &=\tfrac{1}{3!}\big(\{X+\alpha,\mu_2(Y+\beta,Z+\gamma)\}+\mbox{cycl.}\big)~.
\end{aligned}
\end{equation}
The explicit form of~\eqref{eq:ass_Courant_algebra} now depends on the chosen coordinate system since ordinary 1-forms and vector fields are no longer global objects. We first note, that in the holonomic frame $(\tilde \xi^i,\tilde \zeta_i)$, the expressions are just the ones given in~\cite{Deser:2016qkw}, i.e.
\begin{equation}\label{eq:ass_Courant_algebra_holonomic}
\begin{aligned}
 \mu_1(f)&= \dd f~,\\
 \mu_2(X+\alpha,Y+\beta)&= [X,Y] + \CL_X \beta - \CL_Y \alpha  - \frac{1}{2}\dd(\iota_X\beta - \iota_Y \alpha) + \iota_X \iota_Y H~,\\
 \mu_2(X+\alpha,f)&= \CL_X f ~,\\
 \mu_3(X+\alpha,Y+\beta,Z+\gamma)&=\tfrac{1}{3!}\big(\iota_X \iota_Y \dd\gamma + \tfrac{3}{2} \iota_X \dd\iota_Y \gamma \pm \textrm{perm.}\big) + \iota_X\iota_Y\iota_Z H~.
\end{aligned}
\end{equation}

In the nonholonomic frame $(\xi^i,\zeta_i)$, the expressions look formally the same if we use the vielbein. The homological function takes the form $\CQ = \xi^i\tau_i ^j p_j+k\xi^1\xi^2\xi^3$, so we get a new bracket $[\cdot,\cdot]^\tau$ and differential $\dd^\tau$ given for vector fields $X=X^i\zeta_i$, $Y=Y^i\zeta_i$ and forms $\omega=\tfrac{1}{k!}\omega_{i_1\dots i_k}\xi^{i_1}\dots \xi^{i_k}$ by
\begin{equation}
  \begin{aligned}
    [X,Y]^\tau =& (X^i \tau_i^j \partial_j Y^k - Y^i\tau_i^j \partial_j X^k)\zeta_k~,\\
    \dd^\tau\omega =& \tfrac{1}{k!}\,\tau^j_{i_0}\,\partial_j \omega_{i_1\dots i_k}\,\xi^{i_0}\dots\xi^{i_k}~,
  \end{aligned}
\end{equation}
where $\dpar_i$ is short for $\der{x^i}$ and a totally antisymmetrized sum is taken over the indices $i_0\dots i_k$ in the last expression. Using $\dd^\tau$ and $[\cdot,\cdot]^\tau$ together with the $H$-flux written in the coordinates $\xi^i$ we arrive at the expressions
\begin{equation}\label{eq:ass_Courant_algebra_nonholonomic}
\begin{aligned}
 \mu_1(f)&= \dd^\tau f~,\\
 \mu_2(X+\alpha,Y+\beta)&= [X,Y]^\tau + \CL^\tau_X \beta - \CL^\tau_Y \alpha  - \frac{1}{2}\dd^\tau(\iota_X\beta - \iota_Y \alpha) + \iota_X \iota_Y H~,\\
 \mu_2(X+\alpha,f)&= \CL^\tau_X f ~,\\
 \mu_3(X+\alpha,Y+\beta,Z+\gamma)&=\tfrac{1}{3!}\big(\iota_X \iota_Y \dd^\tau\gamma + \tfrac{3}{2} \iota_X \dd^\tau\iota_Y \gamma \pm \textrm{perm.}\big) + \iota_X\iota_Y\iota_Z H~.
\end{aligned}
\end{equation}

For our purposes, it will be useful to generalize the underlying graded vector space of $\sL$ from global to local functions. Given a cover $\sqcup_a U_A\twoheadrightarrow N_j$, this corresponds to extending $\sL$ to $\sL^{\rm loc}= \sL^{\rm loc}_{-1}\oplus \sL^{\rm loc}_0$ with 
\begin{equation}
 \sL^{\rm loc}_{-1}=\sqcup_a \CC^\infty(U_a)\eand \sL^{\rm loc}_0=\sqcup_a\CC^\infty_1(\CCC_{j,k}|_{U_a})\cong \sqcup_a\Gamma(E|_{U_a})~.
\end{equation}
Since the brackets only involve local expressions (i.e.~expressions with finitely many derivatives), the higher products $\mu_i$ are not affected by this extension. Since the brackets are given by tensorial expressions, the brackets will be compatible with the gluing conditions.

\subsection{Action of the associated Lie algebra and its action Lie algebra}\label{ssec:action_Lie_algebra_GG}

As explained in~\cite{Deser:2016qkw}, there is a well-defined general notion of an action of an $L_\infty$-algebra on a graded vector space and associated $L_\infty$-algebras of symplectic $L_\infty$-algebroids naturally come with such an action.

In our case, the Lie 2-algebras $\sL$ and $\sL^{\rm loc}$ act on the functions and the local functions on the Courant algebroid $\CCC_{j,k}$ via the Dorfman bracket\footnote{We note that the Courant and Dorfman brackets on nilmanifolds were also given, e.g., in~\cite{Chatzistavrakidis:2013wra}. The N$Q$-manifold picture goes beyond this since it provides also the full Lie 2-algebra of symmetries.},
\begin{equation}\label{eq:Dorman_explicit}
  \begin{aligned}
    \nu_2(X+\alpha,Y+\beta) =& \bigl\lbrace \lbrace \CQ, X+\alpha\rbrace,Y+\beta\bigr\rbrace\,\\
    =&X^i\tau_i^j\partial_j(Y+\beta) + \bar \xi^i \tau_i^j(\partial_j X^j)\beta_j -Y^k\tau_k^i\partial_i (X+\alpha)+ Y^k \bar \xi^i \tau_i^j \partial_j \alpha_k\\
    =& [X,Y]^\tau + \iota_X \dd^\tau \beta + \dd^\tau (\iota_X\beta) - \iota_Y \dd^\tau\alpha~,
  \end{aligned}
\end{equation}
where $X+\alpha,Y+\beta\in \CCC^\infty_1(\CCC_{j,k})$. This action is readily extended to an action of $\sL^{\rm loc}$ on local functions. We can, in fact, regard the Dorfman bracket as a generalized Lie derivative
\begin{equation}\label{eq:gen_Lie_derivative}
 \hat \CL_{X+\alpha} (Y+\beta)=\bigl\lbrace \lbrace \CQ, X+\alpha\rbrace,Y+\beta\bigr\rbrace~,
\end{equation}
as is customary in Generalized Geometry and Double Field Theory. We note that this action together with the commutator 
\begin{equation}\label{eq:commutator_action_GG}
\begin{aligned}
 [\hat \CL_{X_1+\alpha_1},\hat \CL_{X_2+\alpha_2}] (Y+\beta)&=\hat \CL_{X_1+\alpha_1}\hat \CL_{X_2+\alpha_2} Y-\hat \CL_{X_2+\alpha_2}\hat \CL_{X_1+\alpha_1} (Y+\beta)\\
 &=\hat \CL_{\mu_2(X_1+\alpha_1,X_2+\alpha_2)} (Y+\beta)
\end{aligned}
\end{equation}
forms an ordinary Lie algebra\footnote{The Jacobi identity is trivially satisfied for commutators.} $\sLie(\sL^{\rm loc},\nu_2)$ with ``structure constants'' induced by the Courant bracket $\mu_2$. In particular, the action of the Jacobiator of the Lie 2-algebra is trivial.\footnote{This is obvious from a more abstract point of view: the image of $\mu_1$ describes the action of gauge transformations between gauge transformations (level 1) on gauge transformations (level 0). The former always leave the gauge fields invariant, see e.g.~the explanation of the BRST complex in~\cite{Jurco:2018sby}.}

We can readily identify this Lie algebra as a Lie subalgebra of the symplectomorphisms on $\CCC_{j,k}$, since the generalized Lie derivative~\eqref{eq:gen_Lie_derivative} is simply the action of a Hamiltonian vector field with Hamiltonian $\{\CQ,X+\alpha\}$ on the function $Y+\beta$. Moreover, we can restrict ourselves to the degree-preserving symplectomorphisms on $\CCC_{j,k}$. Such symplectomorphisms have been discussed, e.g., in~\cite{Roytenberg:0203110}, and they have general Hamiltonian of degree~2,
\begin{equation}
 \CH=\CH^i  p_i+\tfrac12 \CH_{ij} \xi^i \xi^j+\CH_i{}^j  \xi^i \zeta_j+\tfrac12 \CH^{ij}\zeta_i\zeta_j~.
\end{equation}
The case at hand, $\CH=\{\CQ,X+\alpha\}$, corresponds to
\begin{equation}
 \CH^i=X^i~,~~~\CH_{ij}=\dpar_i\alpha_j-\dpar_j\alpha_i~,~~~\CH_i{}^j=\dpar_iX^j\eand\CH^{ij}=0~.
\end{equation}

To be more specific about the structure of the Lie algebra, let us consider the explicit form of~\eqref{eq:commutator_action_GG}, first for $H=0$:\begin{equation}
\begin{aligned}
 \hat \CL_{\mu_2(X_1+\alpha_1,X_2+\alpha_2)} (Y+\beta)&=[[X_1,X_2],Y]+\iota_{[X_1,X_2]}\dd^\tau\beta+\dd^\tau\iota_{[X_1,X_2]}\beta-\\
&\hspace{1cm}-\iota_Y\dd^\tau(\CL_{X_1}^\tau\alpha_2-\CL^\tau_{X_2}\alpha_1)~.
\end{aligned}
\end{equation}
The Lie algebra of actions of $\hat \CL_X$ is thus clearly
\begin{equation}\label{eq:Lie_algebra_GG}
 \sLie(\sL^{\rm loc},\nu_2)= \mathfrak{diff}(N_j)\ltimes \Omega^2_{\rm closed}(N_j)~,
\end{equation}
where the semidirect product is given by the natural action of the infinitesimal diffeomorphisms on 2-forms. Note that the $\dd^\tau$-closed forms are locally exact and give rise to the local 1-forms which correspond to part of the local functions $\alpha_{1,2}$ of degree~1 on $\CCC_{j,k}$. 

To understand the case $H\neq 0$, we recall the observation made around~\eqref{eq:eB-coord_trafo_gtb} where we replaced the local sections $Y+\beta$ by the appropriate $B$-field corrected expression
\begin{equation}\label{eq:B-field_correction_GG}
\begin{aligned}
 \widetilde{Y+\beta}&:=\de^B(Y+\beta)=Y+\beta+\iota_Y B\\
 &\phantom{:}=\de^{\{-,B\}}(Y+\beta)=Y+\beta+\{Y,B\}~,
\end{aligned}
\end{equation}
where we used $B=\tfrac12 B_{\mu\nu}\xi^\mu\xi^\nu$ in the Poisson brackets. Note that also
\begin{equation}
\{\widetilde{X+\alpha},\widetilde{Y+\beta}\}=\{X+\alpha,Y+\beta\}~.
\end{equation}
The modification of $\CQ$ due to a non-trivial $B$-field is merely its covariantization with respect to the $\de^B$ action. Denoting the $B$-field modified Hamiltonian by $\tilde \CQ$ we have
\begin{equation}
 \tilde \CQ=\CQ+\dd B=\CQ+\{\CQ,B\}=\de^{\{-,B\}}\CQ
\end{equation}
as well as
\begin{equation}
\begin{aligned}
\{\tilde \CQ,\widetilde{Y+\beta}\}&=\{\de^{\{-,B\}}\CQ\,,\,\de^{\{-,B\}}(Y+\beta)\}\\
&=\{\CQ+\{\CQ,B\}\,,\,Y+\beta+\{Y,B\}\}\\
 &=\{\CQ,Y+\beta\}+\{\CQ,\{Y,B\}\}+\{\{\CQ,B\},Y\}\\
 &=\{\CQ,Y+\beta\}+\{\{\CQ,Y\},B\}\\
 &=\de^{\{-,B\}}\{\CQ,Y+\beta\}\\
 &=\widetilde{\{\CQ,Y+\beta\}}
\end{aligned}
\end{equation}
and similarly
\begin{equation}
 \widetilde{\{\{Q,X+\alpha\},Y+\beta\}}=\{\widetilde{\{\CQ,X+\alpha\}},\widetilde{Y+\beta}\}
\end{equation}
Translating from the action on $Y$ to the action on $\de^B Y$ therefore removes the additional terms containing $B$ in the Lie 2-algebra and its action. We thus recover again~\eqref{eq:Lie_algebra_GG}.

\subsection{Finite symmetries of a gerbe}

The symmetries of the gerbe\footnote{More precisely, the symmetries of the gerbe over local patches, which is all that is needed for T-duality. On double overlaps, we also have gerbe isomorphisms parameterized by \v Cech 1-cochains.} $\CCG_{j,k}$ are given by diffeomorphisms on the base manifold $N_j$ as well as gauge transformations. The latter are parameterized by local 1-forms $\alpha$, but only their differential acts non-trivial, shifting the 2-form connective structure 
\begin{equation}
 B\rightarrow \tilde B=B+\delta B\ewith \delta B=\dd^\tau \alpha=\{\CQ,\alpha\}~.
\end{equation}
Therefore, there are gauge transformations between gauge transformations, parameterized by functions $f$ on $N_j$, and changing the 1-forms $\alpha$ according to
\begin{equation}
 \alpha\rightarrow \tilde \alpha=\alpha+\delta \alpha\ewith \delta\alpha=\dd^\tau f=\{\CQ,f\}~.
\end{equation}
The graded vector space of both these transformations arranges now into the categorified Lie algebra $\sL^{\rm loc}$. 

On the Courant algebroid $\CCC_{j,k}$ whose \v Severa class is the Dixmier--Douady class of the gerbe, this associated Lie 2-algebra also acts on graded functions via the Dorfman bracket $\nu_2$. This is relevant, as a subset of these functions can be interpreted as extended vector and polyvector fields (as formalized, e.g., in~\cite{Deser:2016qkw}). Together with the commutator, these actions form the Lie algebra~\eqref{eq:Lie_algebra_GG}, which is readily integrated to the group
\begin{equation}
 \sDiff(N_j)\ltimes \Omega^2_{\rm closed}(N_{j})~.
\end{equation}
This is also the symmetry group of the Courant algebroid $\CCC_{j,k}$, the automorphism group of the underlying bundle endowed with the Courant bracket. We note that the fact that the symmetries of a general Courant bracket are diffeomorphisms and closed 2-forms on the base manifold of the corresponding Courant bracket is well-known, cf.~e.g.~\cite{Gualtieri:2003dx}.

\section{Global Double Field Theory}

\subsection{Topological T-duality}

Performing a T-duality along the fiber direction of the principal circle bundle $N_j$ maps the gerbe $\CCG_{j,k}$ to the gerbe $\CCG_{k,j}$ over the nilmanifold $N_k$. That is, the T-duality interchanges the Chern class $j$ of the principal circle bundle $N_j$ with the Dixmier--Douady class $k$ of the gerbe $\CCG_{j,k}$ over $N_j$. 

A mathematical picture behind this {\em topological T-duality} has been given long ago in~\cite{Bouwknegt:2003vb}, and the example of the gerbes $\CCG_{j,k}$ was discussed in full detail. Let us briefly summarize the essential points important for our purposes.

Consider a circle bundle $\pi:P\rightarrow M$ and first Chern class $c_1(P)=[F]$ together with a gerbe $\CCG$ on $P$ with Dixmier--Douady class $dd(\CCG)=[H]$. Then we have a closed integral 2-form $\hat F$ on $M$ from the integral along the fiber,
\begin{equation}\label{fiberint}
 \hat F(x)=\pi_*(H)(x):=\oint_{\pi^{-1}(x)} H
\end{equation}
and $[\hat F]$ is the first Chern class of a principal circle bundle $\hat \pi:\hat P\rightarrow M$. One can now argue that there is a 3-form $\hat H$ on $\hat P$ with
\begin{equation}
 \hat \pi_*(\hat H)=c_1(P)~,
\end{equation}
which defines the Dixmier--Douady class of a gerbe $\hat \CCG$ on $\hat P$. Altogether, we obtain the following fibration of spaces:
\begin{equation}\label{eq:generic_diagram}
 \myxymatrix{
     & & P\times_{M}\hat P \ar@{->}[dl]_{\pr} \ar@{->}[dr]^{\hat \pr}& & \\
     \CCG \ar@{->}[r] & P\ar@{->}[dr]_{\pi} & & \hat P\ar@{->}[dl]^{\hat \pi} & \hat \CCG\ar@{->}[l] \\
     & & M & &
    }
\end{equation}
Here, $P\times_{M}\hat P$ is the {\em correspondence space} 
\begin{equation}
 P\times_{M}\hat P:=\{(x,\hat x)\in P\times \hat P\,|\, p(x)=\hat p(\hat x) \}~.
\end{equation}
The correspondence space can either be seen as a circle bundle over $P$ with Chern class $\pi^*(c_1(\hat P))$ or as a circle bundle over $\hat P$ with Chern class $\hat \pi^*(c_1(P))$. Given normalized connection 1-forms $A$ and $\hat A$ on $P$ and $\hat P$, by which we mean $\pi_*(A)=\hat \pi_*(\hat A)=1$, we find that 
\begin{equation}\label{dualitycond}
 \hat \pr^*\hat H-\pr^* H=\dd(\pr^* A\wedge \hat \pr^* \hat A) =: \dd \CF~,
\end{equation}
where $\CF$ is a {\em global} 2-form on $P\times_M P$. Note that the latter implies that the gerbe $\pr^*\CCG\otimes \overline{\hat{\pr}^* \hat \CCG}$ is trivial. Using $\CF$, the \emph{T-duality} map was defined in~\cite{Bouwknegt:2003vb} to be the homomorphism $\hat \pr_* \circ e^{\CF \wedge}\circ \pr^* $ where $\hat \pr_* : P\times_M \hat P \rightarrow \hat P$ is defined analogously to~\eqref{fiberint}. It was shown that this map descends to an isomorphism of twisted cohomologies on $(P,H)$ and $(\hat P, \hat H)$, respectively. One of the aims of the following sections is to reformulate this in our language in order to have a starting point to investigate T-duality in later works. 

\subsection{Symplectic pre-N\texorpdfstring{$Q$}{Q}-manifold on the correspondence space}

Let us now return to our example of pairs of nilmanifolds. Here, topological T-duality interchanges the Dixmier--Douady class $k$ of the gerbe $\CCG_{j,k}$ over $N_j$ with the first Chern class $j$ of the nilmanifold $N_j$. We thus specialize and extend the diagram~\eqref{eq:generic_diagram} to
 \vspace*{0.4cm}
\begin{equation}\label{eq:diagram_goal}
 \myxymatrix{
     & \CCG_K\ar@{->}[dr] \ar@{<.>}@/^4ex/[rr]& \CCE_K= \CCE_{k,j}\ar@{->}[d] & \CCC_K \ar@{->}[dl] \ar@{->}[l] & & \\
     \CCC_{j,k}\ar@{<.>}[d] \ar@{->}[dr] & & K:=N_j\times_{T^2}N_k \ar@{->}[dl]^{\pr_1} \ar@{->}[dr]_{\pr_2}& & \CCC_{k,j} \ar@{->}[dl]\ar@{<.>}[d]\\
     \CCG_{j,k} \ar@{->}[r] & N_j\ar@{->}[dr]_{\pi_j} & & N_k\ar@{->}[dl]^{\pi_k} & \CCG_{k,j} \ar@{->}[l] \\
     & & T^2 & &
    }
\end{equation}
where we also added the Courant algebroids $\CCC_{j,k}$ and $\CCC_{k,j}$ to the picture which describe the symmetries of the gerbes $\CCG_{j,k}$ and $\CCG_{k,j}$. Note that the latter two gerbes can be pulled back to $K$ and tensored there to the gerbe 
\begin{equation}
 \CCG_K:=\pr_1^*\CCG_{j,k}\otimes\pr_2^*\CCG_{k,j}~.
\end{equation}
Restricting this gerbe to $N_j\embd K$ and $N_k\embd K$ yields the gerbes $\CCG_{j,k}$ and $\CCG_{k,j}$, respectively. The gerbe $\CCG_K$ is thus indeed a suitable correspondence object on $K$. Its symmetries are described by the Courant algebroid $\CCC_{\CCG_K}$ on $K$ whose \v Severa class is the Dixmier--Douady class of $\CCG_K$. 

As explained in the introduction, our goal is to characterize the space $\CCE_K$, the appropriate correspondence object on $K$ for Double Field Theory. Since $\CCE_K$ captures the symmetries and the differential geometry of the Double Field Theory relevant for the T-duality between $\CCG_{j,k}$ and $\CCG_{k,j}$, we expect it to be a subspace of $\CCC_{\CCG_K}$. 

This point of view is further supported by our general discussion of the local case in~\cite{Deser:2016qkw}, which also suggests that $\CCE_K$ should be a {\em symplectic pre-N$Q$-manifold of degree~2}. By this we mean a symplectic $\NN$-graded manifold of degree~2 endowed with a vector field $Q$ of degree~$1$, which satisfies $\CL_Q \omega=0$. In particular, a symplectic pre-N$Q$-manifold with $Q^2=0$ is a symplectic N$Q$-manifold. By an {\em $L_\infty$-structure} $\sL$ on a symplectic pre-N$Q$-manifold $\CCE$, we mean a subset of the algebra of functions $\sL\subset \CC^\infty(\CCE)$, such that the usual construction of the associated Lie $n$-algebra on $\CCE$ closes and forms an $L_\infty$-algebra. Translated to the physics nomenclature, the algebraic condition of having an $L_\infty$-structure amounts essentially to (a slight weakening) of the strong section condition and an explicit $L_\infty$-structure is a solution to the strong section condition.

From our explicit description of $N_j$, we readily derive an equally explicit description of the correspondence space $K:=N_j\times_{T^2}N_k$. The latter is a four-dimensional manifold with coordinates $x^\mu$, $\mu=1,\dots,4$, subject to the identification
\begin{equation}
\begin{gathered}
 (x^1\,,\,x^2\,,\,x^3\,,\,x^4)\sim (x^1\,,\,x^2+1\,,\,x^3\,,\,x^4)\sim (x^1\,,\,x^2\,,\,x^3+1\,,\,x^4)\\
 \hspace{3cm}\sim (x^1\,,\,x^2\,,\,x^3\,,\,x^4+1)\sim (x^1+1\,,\,x^2\,,\,x^3-jx^2\,,\,x^4-kx^2)~.
\end{gathered}
\end{equation}

We now extend $K$ to the Courant algebroid $\CCC_{\CCG_K}$ over this space for the gerbe $\CCG_K$, which is also straightforward, as it combines the Courant algebroids $\CCC_{j,k}$ and $\CCC_{k,j}$. We have coordinates $(x^\mu, \xi^\mu,\zeta_\mu,p_\mu)$, $\mu=1,\dots, 4$ subject to the identification 
\begin{multline}
(x^1\,,\,\xi^1\,,\,\xi^2\,,\,\xi^3\,,\,\xi^4\,,\,\zeta_1\,,\,\zeta_2\,,\,\zeta_3\,,\,\zeta_4\,,\,p_1\,,\,p_2\,,\,p_3\,,\,p_4)\\
\sim~(x^1+1\,,\,\xi^1\,,\,\xi^2\,,\,\xi^3\,,\,\xi^4\,,\,\zeta_1\,,\,\zeta_2\,,\,\zeta_3\,,\,\zeta_4\,,\,p_1\,,\, p_2 + jp_3+kp_4\,,\, p_3\,,\,p_4)~. 
\end{multline}
The symplectic form $\omega$ and the Hamiltonian $\CQ$ of the homological vector field are then
\begin{equation}
 \omega=\dd x^\mu\wedge \dd p_\mu+\dd \xi^\mu\wedge \dd \zeta_\mu~,~~~\CQ=\xi^\mu p_\mu-jx^1\xi^2p_3-kx^1\xi^2p_4+k\,\xi^1\xi^2\xi^3+j\,\xi^1\xi^2\xi^4~.
\end{equation}
We note the expected symmetry under the exchange of $j$ and $k$.

To construct the pre-N$Q$-manifold $\CCE_K$, we now linearly combine the fiber coordinates of degree~1 corresponding to the directions involved in the T-duality to new coordinates $\theta^\alpha_\pm$, $\alpha=3,4$:
\begin{equation}
  \theta^3_\pm :=\frac{1}{2}\Bigl( \xi^3 \pm  \zeta_4\Bigr)~,\quad \theta_\pm ^4 :=\frac{1}{2}\Bigl( \xi^4 \pm  \zeta_3\Bigr)~,
\end{equation}
closely following the construction in the local case as outlined in~\cite{Deser:2016qkw}. The pre-N$Q$-manifold $\CCE_K$ in which we are interested is then obtained as the locus of $\theta^\alpha_-=0$, $\alpha=3,4$. More formally, $\CC^\infty(\CCE_K)=\CC^\infty(\CCC_{\CCG_K})/\CI$, where $\CI$ is the ideal generated by the $\theta^\alpha_-$. 

Restricting the symplectic form and the Hamiltonian $\CQ$ to $\CCE_K$ by pull-back along the implied embedding, we obtain
\begin{equation}
 \begin{aligned}
  \omega &= \dd  x^\mu \wedge \dd  p_\mu + \dd \xi^1 \wedge \dd \zeta_1 + \dd \xi^2 \wedge \dd \zeta_2 + \dd\theta_+^3 \wedge \dd\theta_+^4\\
  &=\dd  x^a \wedge \dd  p_a+\dd x^\alpha\wedge \dd p_\alpha + \dd \xi^a \wedge \dd \zeta_a +\tfrac12 \eta_{\alpha\beta}\dd\theta_+^\alpha \wedge \dd\theta_+^\beta
 \end{aligned}
\end{equation}
and
\begin{equation}\label{QE}
  \begin{aligned}
    \CQ &=  \xi^a  p_a + \theta_+^\alpha  p_\alpha- j x^1 \xi^2  p_3 - k x^1  \xi^2  p_4+k\xi^1\xi^2\theta^3_++j\xi^1\xi^2\theta^4_+~,
  \end{aligned}
\end{equation}
where we split all coordinates as $x^\mu=(x^a,x^\alpha)$, etc., $a=1,2$, $\alpha=3,4$ and used the $\sO(1,1)$-metric
\begin{equation}
 \eta_{\alpha\beta}=\begin{pmatrix} 0 & 1 \\ 1 & 0 \end{pmatrix}_{\alpha\beta}
\end{equation}
with inverse $\eta^{\alpha\beta}=\eta_{\alpha\beta}$. The symplectic form $\omega$ induces the Poisson bracket
\begin{equation}\label{eq:Poisson_bracket2}
\begin{aligned}
 \{f,g\}&:=\left(\der{p_\mu} f\right)\left(\der{x^\mu} g\right)-\left(\der{x^\mu}f\right)\left(\der{p_\mu} g\right)\\
 &\hspace{1cm}-(-1)^{|f|}\left(\der{ \zeta_a}f\right)\left(\der{ \xi^a}g\right)-(-1)^{|f|}\left(\der{ \xi^a}f\right)\left(\der{ \zeta_a} g\right)\\
 &\hspace{1cm}-(-1)^{|f|}\left(\der{\theta^4_+}f\right)\left(\der{\theta^3_+}g\right)-(-1)^{|f|}\left(\der{ \theta^3_+}f\right)\left(\der{\theta^4_+} g\right)~.
\end{aligned}
\end{equation}

\noindent We note that $\omega$ and $\CQ$ render $\CCE_K$ a \emph{pre}-N$Q$-manifold, since $\CL_Q \omega = 0$ for the corresponding hamiltonian vector field, and 
\begin{equation}\label{eq:factorization_Qsq}
 \{\CQ,\CQ\}=2p_3p_4=\eta^{\alpha\beta}p_\alpha p_\beta~.
\end{equation}
This is precisely the relation also found in the local case discussed in~\cite{Deser:2016qkw}. This pre-N$Q$-manifold describes indeed the target space for double field theory manifestly invariant under the T-duality connecting the nilmanifolds $N_j$ and and $N_k$ endowed with gerbes of Dixmier--Douady class $k$ and $j$, respectively, as we shall see in the following.

We also note that the above constructions are readily extended to general pairs of Courant algebroids $(\CCC_{j,k},\CCC_{m,n})$ between which there is no T-duality. The factorization of $\{\CQ,\CQ\}$ would here have read as
\begin{equation}\label{eq:non-T-duality-pair}
 \{\CQ,\CQ\}=2(p_3+(k-m)\xi^1\xi^2)(p_4+(n-j)\xi^1\xi^2)~.
\end{equation}
We see that this factorization simplifies for T-dual pairs.

\subsection{Symmetries, \texorpdfstring{$L_\infty$}{Linfinity}-structures and the section condition}

The symmetries of Double Field Theory on $K=N_j\times_{T^2}N_k$ are captured by an {\em $L_\infty$-structure} on $\CCE_K$ as defined in~\cite{Deser:2016qkw}. Recall that an $L_\infty$-structure on $\CCE_K$ is a subset $\sL^{\rm loc}$ of $\CC^\infty_{\rm loc}(\CCE_K)$ (we restrict immediately to local functions),
\begin{subequations}\label{eq:L_infty_struct}
\begin{equation}
 \sL^{\rm loc}=\sL^{\rm loc}_0\oplus\sL^{\rm loc}_1\subset \CC^\infty_{{\rm loc},1}(\CCE_K)\oplus\CC^\infty_{{\rm loc},0}(\CCE_K)
\end{equation}
which becomes a Lie 2-algebra with the following higher brackets:
\begin{equation}\label{eq:ass_DFT_algebra}
\begin{aligned}
 \mu_1(f)&=\{\CQ,f\}=\xi^\mu \tau_\mu^\nu \dpar_\nu=\dd^\tau f~,\\
 \mu_2(X,Y)&=\tfrac12\big(\{\{\CQ,X\},Y\}-\{\{\CQ,Y\},Z\}\big)=[X,Y]^\tau+\tfrac12(Y^A\dd^\tau X_A-X^A\dd^\tau Y_A)~,\\
 \mu_2(X,f)&=\tfrac12\{\{\CQ,X\},f\}=\tfrac12\{X,\{\CQ,f\}\}=\tfrac12X^\mu \tau_\mu^\nu\dpar_\nu f=\tfrac12\CL^\tau_X f~,\\
 \mu_3(X,Y,Z)&=\tfrac{1}{3!}\Big(\{\{\{\CQ,X\},Y\},Z\}\pm\mbox{perm.}\Big)\\
 &=
 Z_A X^\mu \tau_\mu^\nu\dpar_\nu Y^A+X_A Y^\mu \tau_\mu^\nu\dpar_\nu Z^A+Y_A Z^\mu \tau_\mu^\nu\dpar_\nu X^A~.
\end{aligned}
\end{equation}
\end{subequations}
Here, $X,Y,Z\in \sL^{\rm loc}_0\subset\CC_{{\rm loc},1}^\infty(\CCE_K)$ and $f\in \sL^{\rm loc}_1\subset \CC_{{\rm loc},0}^\infty(\CCE_K)$. We also used the index notation
\begin{subequations}\label{eq:index_notation}
\begin{equation}
\begin{gathered}
  x^\mu=( x^a, x^\alpha)~,~~~ x_\alpha=\eta_{\alpha\beta} x^\beta~,~~~
 \eta_{\alpha\beta}=\begin{pmatrix}
                     0 & 1 \\ 1 & 0
                    \end{pmatrix}~,\\
                    (\xi^a,\zeta_a,\theta^\alpha_+)=(\xi^a,\zeta_\mu)=(\xi^\mu,\zeta_a)=\Xi^A
\end{gathered}
\end{equation}
so that
\begin{equation}
 X_a \xi^a+X^a  \zeta_a+X_\alpha\theta^\alpha_+=X_a \xi^a+X^\mu \zeta_\mu=X_\mu \xi^\mu+X^a\zeta_a=X_A \Xi^A
\end{equation}
and
\begin{equation}
 X_A=\eta_{AB}X^B~,~~~
 \eta=\begin{pmatrix}
       0 & \unit_2 & 0 & 0 \\ \unit_2 & 0 & 0 & 0\\ 0 & 0 & 0 & 1\\ 0 & 0 & 1 & 0
      \end{pmatrix}~.
\end{equation}
The vielbein $\tau_\mu^\nu$ is the four-dimensional generalisation of the vielbein $\tau_i^j$, cf.~\eqref{eq:def_vielbein3d}:
\begin{equation}
 \tau_\mu^\nu=\delta_\mu^\nu-jx^1\delta_\mu^2\delta^\nu_3-kx^1\delta_\mu^2\delta^\nu_4~,
\end{equation}
and gives rise to the modified Lie and total derivatives,
\begin{equation}
 [X,Y]^\tau=X^\mu \tau_\mu^\nu \dpar_\nu Y-Y^\mu \tau_\mu^\nu \dpar_\nu X\eand \dd^\tau=\xi^\mu \tau_\mu^\nu \dpar_\nu~.
\end{equation}
\end{subequations}

As shown in~\cite[Theorem~4.7]{Deser:2016qkw}, the conditions for~\eqref{eq:L_infty_struct} to be an $L_\infty$-structure are that the brackets and the Poisson bracket close on $\sL^{\rm loc}$ and that further 
\begin{equation}\label{eq:conditions_thm_Lie2_subset}
\begin{aligned}
\{Q^2f,g\}+\{Q^2g,f\}&=0~,\\
\{Q^2X,f\}+\{Q^2f,X\}&=0~,\\
\{\{Q^2X,Y\},Z\}_{[X,Y,Z]}&=0
\end{aligned}
\end{equation}
for all $f,g\in \sL^{\rm loc}_1$ and $X,Y,Z\in \sL^{\rm loc}_0$. 

Explicitly, the elements of $\sL^{\rm loc}_0$ parameterize infinitesimal diffeomorphisms on $K$ as well as the gauge transformations of the $B$-field, which is part of the connective structure of the gerbe $\CCG_K$. The elements of $\sL^{\rm loc}_1$ are higher gauge transformations, that is gauge transformations between gauge transformations which encode the redundancy in the gauge transformations parameterized by $\sL^{\rm loc}_0$.

The algebraic conditions~\eqref{eq:conditions_thm_Lie2_subset} for having an $L_\infty$-structure translate to the section condition in the usual Double Field Theory nomenclature. They are necessary for having an appropriate symmetry algebra on the correspondence space. In~\cite{Deser:2016qkw}, we showed that in the local case, they correspond in fact to a weakening of the section condition.

\subsection{\texorpdfstring{$L_\infty$}{Linfinity}-algebra structures: solutions to the section condition}

Let us now come to discussing the solutions to the section condition which reduce $\CCE_K$ to the Courant algebroids $\CCC_{j,k}$ and $\CCC_{k,j}$. The factorization $\{\CQ,\CQ\}=2p_3p_4$ gives a clear hint that we want to quotient by the Hamiltonian vector fields of $p_3$ and $p_4$,
\begin{equation}
 \der{x^3}=\{p_3,-\}\eand \der{x^4}=\{p_4,-\}~.
\end{equation}
We want to regard these vector fields as part of integrable distributions which encode foliations whose leaves are the fibers to the projections
\begin{equation}\label{eq:projections}
 \CCE_K\xrightarrow{~\hat \pr_1~}\CCC_{j,k}\eand \CCE_K\xrightarrow{~\hat \pr_2~}\CCC_{k,j}~.
\end{equation}
For this, we clearly need an additional pair of vector fields $V_3$ and $V_4$ completing the distributions. It is natural to choose the symplectic dual vector fields to the global 1-forms 
\begin{equation}
 \dd x^3+j x^1\dd x^2 \eand \dd x^4+k x^1\dd x^2~.
\end{equation}
Up to a sign, we then arrive at
\begin{equation}
 V_3=\der{ p_3}+j x^1\der{p_2}\eand V_4=\der{ p_4}+k x^1\der{p_2}~,
\end{equation}
which satisfy
\begin{equation}
 \iota_{V_3}\omega=-(\dd x^3+j x^1\dd x^2)\eand \iota_{V_4}\omega=-(\dd x^4+k x^1\dd x^2)
\end{equation}
and are therefore global\footnote{The symplectic form is global and invertible and so are the two 1-forms.}. We note that both $(\dpar_3,V_3)$ and $(\dpar_4,V_4)$ span integrable distributions, giving indeed rise to the projections $\hat \pr_2$ and $\hat \pr_1$, respectively. Explicitly, we have the obvious projections
\begin{equation}\label{eq:projection}
\begin{aligned}
 &\hat \pr_1( x^1\,,\,x^2\,,\,x^3\,,\,x^4\,,\, \xi^1\,,\,\xi^2\,,\, \zeta_1\,,\, \zeta_2\,,\,\theta^3_+\,,\,\theta^4_+\,,\,p_1\,,\,p_2\,,\,p_3\,,\,p_4)\\
 &~~~=(\tilde{{x}}^1\,,\,\tilde{{x}}^2\,,\,\tilde{{x}}^3\,,\, \tilde{{\xi}}^1\,,\,\tilde{{\xi}}^2\,,\,\tilde{{\xi}}^3\,,\,\tilde{{\zeta}}_1\,,\,\tilde{{\zeta}}_2\,,\,\tilde{{\zeta}}_3\,,\,\tilde{{p}}_1\,,\,\tilde{{p}}_2\,,\,\tilde{{p}}_3)\\
 &~~~=( x^1\,,\, x^2\,,\, x^3\,,\, \xi^1\,,\, \xi^2\,,\,\theta^3_+\,,\, \zeta_1\,,\,\zeta_2\,,\,\theta^4_+\,,\, p_1\,,\, p_2-k  x^1 p_4\,,\, p_3)~\,,\,\\
 &\hat \pr_2( x^1\,,\,x^2\,,\,x^3\,,\,x^4\,,\, \xi^1\,,\,\xi^2\,,\, \zeta_1\,,\, \zeta_2\,,\,\theta^3_+\,,\,\theta^4_+\,,\,p_1\,,\,p_2\,,\,p_3\,,\,p_4)\\
 &~~~=(\tilde{{x}}^1\,,\,\tilde{{x}}^2\,,\,\tilde{{x}}^3\,,\, \tilde{{\xi}}^1\,,\,\tilde{{\xi}}^2\,,\,\tilde{{\xi}}^3\,,\,\tilde{{\zeta}}_1\,,\,\tilde{{\zeta}}_2\,,\,\tilde{{\zeta}}_3\,,\,\tilde{{p}}_1\,,\,\tilde{{p}}_2\,,\,\tilde{{p}}_3)\\
 &~~~=( x^1\,,\, x^2\,,\, x^4\,,\, \xi^1\,,\, \xi^2\,,\,\theta^4_+\,,\, \zeta_1\,,\,\zeta_2\,,\,\theta^3_+\,,\, p_1\,,\, p_2-j  x^1 p_3\,,\, p_4)~\,,\,
\end{aligned}
\end{equation}
where $(\tilde{{x}}^1,\dots,\tilde{{p}}_3)$ are the coordinates on the Courant algebroids $\CCC_{j,k}$ and $\CCC_{k,j}$. Both projections allow for sections, which are the evident embeddings of $\CCC_{j,k}$ and $\CCC_{k,j}$ into the pre-N$Q$-manifold $\CCE_K$:
\begin{equation}\label{embeddings}
\begin{aligned}
 &e_1(\tilde{{x}}^1\,,\,\tilde{{x}}^2\,,\,\tilde{{x}}^3\,,\, \tilde{{\xi}}^1\,,\,\tilde{{\xi}}^2\,,\,\tilde{{\xi}}^3\,,\,\tilde{{\zeta}}_1\,,\,\tilde{{\zeta}}_2\,,\,\tilde{{\zeta}}_3\,,\,\tilde{{p}}_1\,,\,\tilde{{p}}_2\,,\,\tilde{{p}}_3)\\
 &~~~=( x^1\,,\,x^2\,,\,x^3\,,\,x^4\,,\, \xi^1\,,\,\xi^2\,,\, \zeta_1\,,\, \zeta_2\,,\,\theta^3_+\,,\,\theta^4_+\,,\,p_1\,,\,p_2\,,\,p_3\,,\,p_4)\\
 &~~~=(\tilde{{x}}^1\,,\,\tilde{{x}}^2\,,\,\tilde{{x}}^3\,,\,-k\tilde{{x}}^1\tilde{{x}}^2\,,\,\tilde{{\xi}}^1\,,\,\tilde{{\xi}}^2\,,\,\tilde{{\zeta}}_1\,,\,\tilde{{\zeta}}_2\,,\,\tilde{{\xi}}^3\,,\,\tilde{{\zeta}}_3\,,\,\tilde{{p}}_1\,,\,\tilde{{p}}_2\,,\,\tilde{{p}}_3\,,\,0)~\,,\,\\
 &e_2(\tilde{{x}}^1\,,\,\tilde{{x}}^2\,,\,\tilde{{x}}^3\,,\, \tilde{{\xi}}^1\,,\,\tilde{{\xi}}^2\,,\,\tilde{{\xi}}^3\,,\,\tilde{{\zeta}}_1\,,\,\tilde{{\zeta}}_2\,,\,\tilde{{\zeta}}_3\,,\,\tilde{{p}}_1\,,\,\tilde{{p}}_2\,,\,\tilde{{p}}_3)\\
 &~~~=( x^1\,,\,x^2\,,\,x^3\,,\,x^4\,,\, \xi^1\,,\,\xi^2\,,\, \zeta_1\,,\, \zeta_2\,,\,\theta^3_+\,,\,\theta^4_+\,,\,p_1\,,\,p_2\,,\,p_3\,,\,p_4)\\
 &~~~=(\tilde{{x}}^1\,,\,\tilde{{x}}^2\,,\,-j\tilde{{x}}^1\tilde{{x}}^2\,,\,\tilde{{x}}^3\,,\,\tilde{{\xi}}^1\,,\,\tilde{{\xi}}^2\,,\,\tilde{{\zeta}}_1\,,\,\tilde{{\zeta}}_2\,,\,\tilde{{\zeta}}_3\,,\,\tilde{{\xi}}^3\,,\,\tilde{{p}}_1\,,\,\tilde{{p}}_2\,,\,0\,,\,\tilde{{p}}_3)~.
\end{aligned}
\end{equation}
These embeddings are morphisms of symplectic pre-N$Q$-manifolds: the symplectic forms and the Hamiltonians of the homological vector fields on the Courant algebroids are pullbacks of those on $\CCE_K$ along $e_1$ and $e_2$.\footnote{Mathematically, we can thus encode T-duality as a cospan of pre-N$Q$-manifolds, but we refrain from pushing this thought any further in the current paper.}

It is now easy to show that the subsets
\begin{equation}\label{eq:L_infty_algebra_structures}
 \sL_{\hat \pr_1}=\{F \in \CC_{\rm loc}^\infty(\CCE_K)~|~\dpar_4F=V_4F=0\}~~\mbox{and}~~
 \sL_{\hat \pr_2}=\{F \in \CC_{\rm loc}^\infty(\CCE_K)~|~\dpar_3F=V_3F=0\}
\end{equation}
indeed form $\sL_\infty$-algebra structures, verifying conditions~\eqref{eq:conditions_thm_Lie2_subset}. For this, we rewrite
\begin{equation}
 Q^2F=\{\CQ,\{\CQ,F\}\}=\tfrac12 \{\{\CQ,\CQ\},F\}
\end{equation}
for $F\in \CC^\infty_{\rm loc}(\CCE_K)$. We also note that
\begin{equation}
 \{\{\CQ,\CQ\},F\}=2\{p_3p_4,F\}=\begin{cases}
                                 2p_4\{p_3,F\}~\mbox{for}~F\in \sL_{\hat \pr_1}~,\\
                                 2p_3\{p_4,F\}~\mbox{for}~F\in \sL_{\hat \pr_2}~.
                                \end{cases}
\end{equation}

We now restrict our discussion to the case $\sL_{\hat \pr_1}$; the case $\sL_{\hat \pr_2}$ follows fully analogously. We have
\begin{subequations}\label{eq:special_properties_L_1}
\begin{equation}
  \{Q^2f,g\}= p_4\{\{p_3,f\},g\}=0
\end{equation}
for $f,g\in \sL_{\hat \pr_1,1}$, since $\{p_3,f\}$ and $g$ are functions of degree~0 and the Poisson bracket on $\CCE_K$ between two such functions vanishes. Similarly,
\begin{equation}
\begin{aligned}
 \{Q^2X,f\}= p_4\{\{p_3,X\},f\}=0~,\\
 \{Q^2f,X\}=p_4\{\{p_3,f\},X\}=0
\end{aligned}
\end{equation}
for $f\in \sL_{\hat \pr_1,1}$ and $X\in \sL_{\hat \pr_1,0}$ since the Poisson bracket also vanishes between a function of degree~0 and a function of degree~1. Finally, we have
\begin{equation}
 \{\{Q^2 X,Y\},Z\}= p_4\{\{\{p_3,X\},Y\},Z\}=0
\end{equation}
\end{subequations}
for $X,Y,Z\in \sL_{\hat \pr_1,0}$, since $\{\{p_3,X\},Y\}$ is a function of degree~0 on $\CCE_K$. Altogether,~\eqref{eq:special_properties_L_1} are sufficient for equations~\eqref{eq:conditions_thm_Lie2_subset} to be satisfied. Note, however, that they are by no means necessary. They are rather special, in that each term in the symmetrizations contained in~\eqref{eq:conditions_thm_Lie2_subset} vanishes separately.

A few remarks are in order. First, we have not used $V_4 F=0$ for $F\in \CCC^\infty_{\rm loc}(\CCE_K)$ at any point of the proof. Thus the subsets $\{F \in \CC_{\rm loc}^\infty(\CCE_K)~|~\dpar_4F=V_4F=0\}$ and $\{F \in \CC_{\rm loc}^\infty(\CCE_K)~|~\dpar_3F=V_3 F=0\}$ determining the projected Courant algebroids also form $\sL^{\rm loc}_\infty$ algebras, as it is to be expected from the fact that they represent Courant algebroids. The condition $V_4F = 0$ seems to be merely necessary for reducing to a symplectic space in the projection $\hat \pr_1$ of~\eqref{eq:projections} and analogously for $V_3 F = 0$. Second, the above discussion straightforwardly extends to general pairs of Courant algebroids $(\CCC_{j,k},\CCC_{m,n})$ between which there is no T-duality. The factorization~\eqref{eq:non-T-duality-pair} of $\{\CQ,\CQ\}$ yields two different vector fields,
\begin{equation}
D_3=\{p_3+(k-m)\xi^1\xi^2,-\}\eand D_4=\{p_4+(n-j)\xi^1\xi^2,-\}~.
\end{equation}
The projection map is correspondingly more complicated.

\subsection{T-duality between \texorpdfstring{$\CCC_{j,k}$}{Cjk} and \texorpdfstring{$\CCC_{k,j}$}{Cjk}}

We finally want to investigate the meaning of T-duality between $(N_j,H)$ and $(N_k,\hat H)$ in our language. Recalling diagram~\eqref{eq:diagram_goal}, we first rewrite the condition~\eqref{dualitycond} in our language. Noting that the differential $\dd$ in~\eqref{dualitycond} is the untwisted de Rham differential on the correspondence space $K$, we split the homological function~\eqref{QE} according the untwisted part plus the flux part:
\begin{equation}
  \CQ = \CQ_0 + k\xi^1\xi^2\theta_+^3 + j\xi^1\xi^2\theta_+^4\;.
\end{equation}
We further note that the product of the connection forms, i.e.~$\pr^* A \wedge \hat \pr^* \hat A$, lifted to $\CE_K$ takes the simple form $\CF = -\theta^3_+ \theta^4_+$. Hence, equation~\eqref{dualitycond} takes the form
\begin{equation}\label{NQcond}
  \lbrace \CQ_0, \CF\rbrace = -k\xi^1\xi^2\theta_+^3 + j\xi^1\xi^2\theta_+^4\;.
\end{equation}
Using this condition, it turns out that we are able to transport the homological function on $\CCC_{j,k}$ to its corresponding object on $\CCC_{k,j}$ using the projections $\hat \pr_i$ of~\eqref{eq:projection} and embeddings in~\eqref{embeddings} in composition:
\begin{equation}
 \CQ_{k,j} = e_2^* \circ \hat \pr_1^* \CQ_{j,k}~,
\end{equation}
where $\CQ_{j,k}$ and $\CQ_{k,j}$ are the homological functions on $\CCC_{j,k}$ and $\CCC_{k,j}$, respectively. Evaluating the left hand side of~\eqref{NQcond} gives the relation for the twists:
\begin{equation}\label{intermediate}
\theta_+^3(p_3 + k\xi^1\xi^2) = \theta_+^4(p_4 + j\xi^1\xi^2)\;.
\end{equation}
This enables the exchange of the parts differing in the pullbacks of the homological functions:
\begin{equation}
\begin{aligned}
    \hat \pr_1^* \CQ_{j,k} =& \xi^1 p_1 + \xi^2 p_2 + \theta_+ ^3 p_3 - \xi^2 x^1(k p_4 + jp_3) + k\xi^1\xi^2 \theta_+^3\nonumber\\
    =& \xi^1 p_1 + \xi^2 p_2 + \theta^4_+ p_4 - \xi^2 x^1(kp_4 + jp_3) + j\xi^1 \xi^2\theta_+^4 \nonumber
    =& \hat \pr_2 ^* \CQ_{k,j}\;.
\end{aligned}
\end{equation}
It follows from the definition of $e_2$ that $e_2^* \circ \hat \pr_1 ^* \CQ_{j,k} = \CQ_{k,j}$. We also have $e_2^*\circ \hat \pr_1^* H = k\tilde \xi^1\tilde \xi^2 \tilde \zeta_3$, which is the ``$f$-flux'' T-dual to the 3-form flux $H$.
 
We see that there are two steps for T-duality here: First, the factorization of~\eqref{eq:factorization_Qsq} determines projections $\hat \pr_i$ from the pre-$NQ$-manifold $\CE_K$ to the respective Courant algebroids. And second, the condition~\eqref{dualitycond} allows to relate the homological functions on the N$Q$-manifolds corresponding to the projected Courant algebroids. This is the reformulation of the T-duality map in our language. Of course we restricted to the very example of the two nilmanifolds with interchanged Chern- and Dixmier--Douady classes, but we expect this to hold true whenever geometric T-duality is possible.

\subsection{\mathversion{bold}Differential geometry on \texorpdfstring{$\CCE_K$}{EK}}

Let us briefly comment also on the differential geometry on $\CCE_K$. The very general framework developed in~\cite{Deser:2016qkw} is readily applied to the case at hand. 

The pre-N$Q$-manifold $\CCE_K$ by itself describes the symmetries of Double Field Theory, which act on extended tensors. For these, we need to extend the algebra of functions on $\CCE_K$ to the free tensor algebra $T(\CCE_K)$ generated by the local coordinate functions over the ring $\CC^\infty_{{\rm loc},0}(\CCE_K)$. The Poisson bracket now uniquely extends to a map
\begin{equation}
 \{-,-\}: \CC^\infty_{\rm loc}(\CCE_K)\times T(\CCE_K)\rightarrow T(\CCE_K)~,
\end{equation}
which encodes in particular the action of elements of an $L_\infty$-structure $\sL^{\rm loc}$ on $\CCE_K$ on a subset $\sT(\CCE_K)\subset T(\CCE_K)$. In order for this action to be a reasonable action of a 2-term $L_\infty$-algebra on a vector space, we require that
\begin{equation}\label{eq:tensor-conditions}
\big\{\{Q^2X,Y\}-\{Q^2Y,X\},t\big\}=0\eand \Big\{\big\{\{Q^2X,Y\}-\{Q^2Y,X\},QZ\big\},t\Big\}=0
\end{equation}
for all $t\in \sT(\CCE_K)$ as well as all $X,Y,Z\in \sL^{\rm loc}_0$, cf.~\cite[Theorem~4.11]{Deser:2016qkw}. This condition is the analogue for tensors of the condition~\eqref{eq:conditions_thm_Lie2_subset} for the parameters of infinitesimal symmetries. We note that both conditions are trivially satisfied if we consider functions $F$ (and the obvious extension to elements of the tensor algebra) with $\dpar_4 F=V_4 F=0$ or $\dpar_3 F=V_3 F=0$. So the above extended tensors on $\CCE_K$ reduce to the extended tensors on the Courant algebroids $\CCC_{j,k}$ and $\CCC_{k,j}$.

Our formalism~\cite{Deser:2016qkw} also knows extended notions of covariant derivative, torsion and Riemann tensor. These merely rely on the available pre-N$Q$-structures. An extended covariant derivative $\nabla$ on $\CCE_K$ is defined as a linear map from $\CCX(\CCE_K)$ to $\CC^\infty(\CCE_K)$ such that the image $\nabla_X$ for $X\in \CCX(\CCE_K)$ gives rise to a map
\begin{equation}
\{\nabla_X,-\}:\CCX(\CCE_K)\rightarrow \CCX(\CCE_K)~,
\end{equation}
which readily generalizes to extended tensors and satisfies 
\begin{equation}
\{\nabla_{fX},Y\}=f\{\nabla_X,Y\}\eand\{\nabla_X,fY\}=\{Q X,f\}Y+f\{\nabla_X,Y \}
\end{equation}
for all $f\in \CCC_0^\infty(\CCE_K)$ and extended tensors $Y\in \sT(\CCE_K)$. For this notion of extended covariant derivative, we readily write down the extended torsion and extended Riemann tensors,
\begin{equation}
 \begin{aligned}
  &\CT : \otimes^3 \CCX(\CCE_K) \rightarrow \CC^\infty_0(\CCE_K)~,\\
  &\CT(X,Y,Z):=\, 3\Bigl(-\Bigl\{X,\{\nabla_{Y},Z\}\Bigr\}\Bigr)_{[X,Y,Z]}+\frac{1}{2}\left(\{X,\{QZ,Y\}\}-\{Z,\{QX,Y\}\}\right)~,\\
  &\CR : \otimes^4 \CCX(\CCE_K) \rightarrow \CC^\infty_0(\CCE_K)~,\\
  &\CR(X,Y,Z,W):=\frac{1}{2}\Bigl(\Bigl\{\bigl\{\{\nabla_X,\nabla_Y\} -\nabla_{\mu_2(X,Y)},Z\bigr\},W\Bigr\} +( Z\leftrightarrow W)   \\
&\hspace{3cm} + \Bigl\{\bigl\{\{\nabla_Z,\nabla_W\}-\nabla_{\{\nabla_Z,W\} - \{\nabla_W,Z\}},X\bigr\},Y\Bigr\} +(X\leftrightarrow Y)\Bigr)~,
 \end{aligned}
\end{equation}
where $X,Y,Z,W\in \CCX(\CCE_K)$, the subset of $\sT(\CCE_K)$ of degree~1. Note that both $\CT$ and $\CR$ are indeed tensorial up to terms which vanish on $L_\infty$-structures.

\section{Finite symmetries of Double Field Theory}

We shall first discuss the finite symmetries of local Double Field Theory in a general context before specializing to the case of the symplectic pre-N$Q$-manifold $\CCE_K$.

\subsection{Generalized Lie derivative and infinitesimal symmetries}

We start from the description of local Double Field Theory as given in~\cite{Deser:2016qkw}. That is, we consider the symplectic pre-N$Q$-manifold $\CCE_2(\FR^d)$, which has local coordinates $x^M,\theta^M,p_M$ of degrees $0$, $1$ and $2$ and index $M=1,\dots,2d$. The symplectic form and the Hamiltonian $\CQ$ read as 
\begin{equation}\label{eq:DFT_gen_symplectic}
 \omega=\dd x^M\wedge \dd p_M+\tfrac12 \eta_{MN}\dd \theta^M\wedge \dd \theta^N\eand \CQ=\theta^M p_M~,
\end{equation}
where $\eta_{MN}$ is the usual $\sO(d,d)$-structure. The Lie 2-algebra structure is given by the same derived bracket as in~\eqref{eq:ass_DFT_algebra}, and $L_\infty$-structures have to satisfy the constraint (or section condition)~\eqref{eq:conditions_thm_Lie2_subset}.

The generalized Lie derivative of Double Field Theory is simply the action of the extended vector fields in the Lie 2-algebra of symmetries on extended vector fields and extended tensors. On extended vector fields, it is given by the D-bracket $\nu_2$, which is the following derived bracket:
\begin{equation}
\begin{aligned}
 \hat \CL_X Y&=\nu_2(X,Y)=\{\{\CQ,X\},Y\}=\left\{X^Mp_M+\theta^M\theta^N\der{x^M} X_N,Y\right\}\\
 &=[X,Y]+\theta^M Y^K\der{x^M} X_K
\end{aligned} 
\end{equation}
for generalized vectors $X=\theta^M X_M$, $Y=\theta^M Y_M$ in $\sL^{\rm loc}_0\subset \CCC^\infty_1(\CCE_2)$. We now want to impose two further restrictions on our $L_\infty$-structure $\sL^{\rm loc}$ and set of extended tensors $\sT$. First of all, we want that symmetries between symmetries have no effect on the action. This implies that
\begin{equation}\label{eq:new_section_1}
 \nu_2(\mu_1(f),Y)=\{\{\CQ,\{\CQ,f\}\},Y\}=0
\end{equation}
for all $f\in \sL^{\rm loc}_{1}\subset \CCC^\infty_0(\CCE_K)$ and $Y\in \sT$. Note that
\begin{equation}
\begin{aligned}
\{\{\CQ,\{\CQ,f\}\},Y\}&=\{\{\tfrac12\{\CQ,\CQ\},f\},Y\}=\tfrac12 \{\{(p_M p^M,f\},Y\}=\left(\der{x^M} f\right)\left(\der{x_M} Y\right)~,
\end{aligned}
\end{equation}
which is the usual section condition on $\CCE_2(M)$, trivially satisfied, for example, for the trivial $L_\infty$-structure 
\begin{equation}
 \sL^{\rm loc}_+=\left\{F\in \CCC^\infty(\CE_2(\FR^d)) ~\big|~ \der{x_\mu} F=\der{p^\mu} F=0\right\}~.
\end{equation}

Second, we would like $\nu_2$ to form a Leibniz algebra on the degree~0 elements $\sL^{\rm loc}_0$ of an $L_\infty$-structure $\sL^{\rm loc}$. As explained in~\cite{Deser:2016qkw}, this also requires that 
\begin{equation}\label{eq:new_section_2}
 \{\{\{\{\CQ,\CQ\},X\},Y\},Z\}=0
\end{equation}
for all $X,Y,Z\in \sL^{\rm loc}_0$. Explicitly,
\begin{equation}
\begin{aligned}
 \{\{\{\{\CQ,\CQ\},X\},Y\},Z\}&=\{\{\{p_Mp^M,X\},Y\},Z\}\\
 &=\{(\dpar_M X)(\dpar^M Y),Z\}+\{\{\dpar_M X,Y\},\dpar^M Z\}~,
\end{aligned}
\end{equation}
again an analogue of the section condition trivially satisfied for the $L_\infty$-structure~$\sL^{\rm loc}_+$ introduced above.

Now the action of the generalized Lie derivative, together with the commutator, forms a Lie algebra, just as in the case of the Courant algebroid. The structure constants are encoded in the C-bracket, i.e.~the higher product $\mu_2$,
\begin{equation}\label{eq:commutator_action_DFT}
 [\hat \CL_{X_1},\hat \CL_{X_2}] Y=\hat \CL_{X_1}\hat \CL_{X_2} Y-\hat \CL_{X_2}\hat \CL_{X_1}Y=\hat \CL_{\mu_2(X_1,X_2)}Y~,
\end{equation}
$X_{1,2}\in \sL^{\rm loc}_0\subset \CCC^\infty_1(\CCE_2)$,
which is easily verified using the derived bracket formulation of this equation. Since the action is Hamiltonian and degree preserving, we know that this has to be a Lie subalgebra of the degree preserving symplectomorphisms on $\CE_2(\FR^d)$ as again familiar from the case of the Courant algebroid. Contrary to that case, however, it is a little bit harder to identify this Lie subalgebra.

We note that the generalized Lie derivative acts as the ordinary Lie derivative on functions $f$ of degree $0$:
\begin{equation}
 \hat \CL_X f=\{\{\CQ,X\},f\}=\{X^Mp_M+\theta^M\dpar_M X,f\}=X^M\dpar_M f=\CL_X f~,
\end{equation}
and the Lie algebra of actions on these functions is simply the Lie algebra of actions of infinitesimal diffeomorphisms on the body of $\CCE_2$, which is homomorphic to the Lie algebra of infinitesimal diffeomorphisms on the body of $\CCE_2$. 

On generalized vectors, which are functions $Y$ of degree~1, however, we have an expression differing from $\CL_X$:
\begin{equation}\label{eq:gen_Lie_derivative_on_Y}
\begin{aligned}
 \hat \CL_X Y&=\{\{\CQ,X\},Y\}=\{X^Mp_M+\theta^M\dpar_M X,Y\}\\
 &=X^M\dpar_M Y-Y^M\dpar_M X+\theta^M Y_K\dpar_M X^K\\
 &=[X,Y]+\theta^M Y_K \dpar_M X^K~.
\end{aligned}
\end{equation}

To find the Lie algebra of general actions, we consider the Leibniz algebra formed by $\nu_2$, which is fully determined by its action on the coordinate functions. We have 
\begin{equation}\label{eq:inf_Hamiltonian_action}
 \begin{aligned}
  \hat \CL_X x^M&=\{\{\CQ,X\}, x^M\}=X^M~,\\
  \hat \CL_X \theta^M&=\{\{\CQ,X\},\theta^M\}=\theta^N\der{x^N} X^M-\der{x_M}X~,\\
  \hat \CL_X p_M&=\{\{\CQ,X\},p_M\}=-p_N\der{x^M}X^N-\theta^N\der{x^N}\der{x^M} X~.
 \end{aligned}
\end{equation}
A similar action was discussed already in~\cite{Roytenberg:0203110}, where also a finite form was given to which we shall return later. The commutator of two actions of the generalized Lie derivative has an interesting form,
\begin{equation}\label{eq:gen_Lie_commutator}
\begin{aligned}
 [\hat \CL_{X_1},\hat \CL_{X_2}]F&=\{\{\CQ,X_1\},\{\{\CQ,X_2\},F\}\}-\{\{\CQ,X_2\},\{\{\CQ,X_1\},F\}\}\\
 &=\{\{\{\CQ,X_1\},\{\CQ,X_2\}\},F\}=\{\{\CQ,\{\{\CQ,X_1\},X_2\}\},F\}\\
 &=\{\{\CQ,[X_1,X_2]+\theta^MX_2^K\dpar_M X_{1K}\},F\}\\
 &=\{[X_1,X_2]^Mp_M+\theta^M\dpar_M[X_1,X_2]-\theta^M\theta^N\dpar_N(X_2^K\dpar_MX_{1K}),F\}\\
 &=\CL_{[X_1,X_2]}F+R(X_1,X_2)F~,
\end{aligned}
\end{equation}
where $F$ is an element of an $L_\infty$-structure and thus $|F|\leq 1$. Here, we used the variants of the section condition
\begin{equation}
 \{\{\{\CQ,\CQ\},X_1\},X_2\},F\}=0\eand \{p^M X_2^K \dpar_M X_{1K},F\}=0~,
\end{equation}
as well as the definition of the ordinary Lie derivative $\CL_X=\dd \iota_X+\iota_X\dd$ via the identification
\begin{equation}
 \dd X=\theta^M\{p_M,X\}\eand \iota_{X_1} X_2=\{X_1,X_2\}~.
\end{equation}
Note that $R(X_1,X_2)F$ is a contraction of $F$ with a local and antisymmetric matrix. So we conclude that the action of the generalized Lie derivative forms the Lie algebra of diffeomorphisms whose action on extended tensors is modified by a local $\ao(d,d)$-transformation. Note that the same conclusion was already reached in~\cite{Berman:2014jba}.

\subsection{Finite symmetries}

A more general form of~\eqref{eq:inf_Hamiltonian_action} has already been integrated in~\cite{Roytenberg:0203110} and the finite transformations are given by
\begin{equation}
 \begin{aligned}
  x^M\mapsto \tilde x^M&=\tilde x^M(x)~,\\
  \theta^M\mapsto \tilde \theta^M&=T^M{}_N(x)\theta^N~,\\
  p_M\mapsto \tilde p_M&=\derr{x^N}{\tilde x^M}p_N+\tfrac12 \theta^N\derr{T^K{}_N(x)}{\tilde x^M} \eta_{KL} T^L{}_P(x) \theta^P~,
 \end{aligned}
\end{equation}
where the first map is a diffeomorphism and $T^M{}_N(x)$ is a local $\sO(d,d)$-transformation, because we are integrating an infinitesimal degree-preserving symplectomorphism and $\eta_{MN}$ is part of the symplectic form, cf.~\eqref{eq:DFT_gen_symplectic}. 

From the infinitesimal case, we know that the choice of diffeomorphism, together with the metric $\eta_{MN}$, induces the local $\sO(d,d)$-transformation. A clear description of the action of the generalized Lie derivative is found by recalling the interpretation of the Dorfman bracket from Generalized Geometry,
\begin{equation}\label{eq:Dorfman_GG}
\begin{aligned}
 \hat \CL^0_{X+\alpha}(Y+\beta)&:=\{\{\CQ_0,X+\alpha\},Y+\beta\}\\
 &\phantom{:}=[X,Y]+\CL_X \beta-\iota_Y \dd\alpha~.
\end{aligned}
\end{equation}
for vector fields $X,Y$ and 1-forms $\alpha,\beta$, where $\CQ_0$ is the Hamiltonian of the homological vector field {\em without} any $B$-field corrections (i.e.~the effects of the $B$-field are in the transition functions). The generalized Lie derivative gives an action of the semidirect product of the Lie algebra of diffeomorphisms and local 1-forms on generalized vectors in Generalized Geometry. There is, however, a second action within ordinary differential geometry, which is given by the first two terms in~\eqref{eq:Dorfman_GG}. In this picture, the last  term describes the additional term in $\de^B (Y+\beta)$ due to the gauge transformation
\begin{equation}\label{eq:gauge_trafo_GG}
\begin{aligned}
 B~~\mapsto~~B'&=B+\delta B=B+\dd \alpha+\hat \CL_{X+\alpha} B\\
 &=B+\dd \alpha+\CL_X B\\
 &=B+\{\CQ_0,\alpha\}+\{\{\CQ_0,X+\alpha\},B\}~.
\end{aligned}
\end{equation}
Explicitly, we have
\begin{equation}
\begin{aligned}
 Y+\beta~~\xmapsto{~~\hat \CL^0_X~~}~~Y'+\beta'&=Y+\beta+\hat \CL_{X+\alpha}(Y+\beta)\\
 &=Y+\beta+[X,Y]+\CL_X \beta-\iota_Y \dd \alpha~,
\end{aligned}
\end{equation}
but after a $B$-field coordinate change, $\de^B(Y+\beta)$ is mapped to $\de^{B'}(Y'+\beta')$ as follows:
\begin{equation}
\begin{aligned}
 \de^{B'}(Y'+\beta')&=Y+\beta+\{\{\CQ_0,X+\alpha\},Y+\beta\}+\{Y+\beta+\{\{\CQ_0,X+\alpha\},Y+\beta\},B\}+\\
 &\hspace{1cm}+\{Y+\beta,\{\CQ_0,\alpha\}+\{\{\CQ_0,X+\alpha\},B\}\}\\
&=Y+\beta+\{Y+\beta,B\}+\{\{\CQ_0,X\},Y+\beta\}+\{\{\CQ_0,X+\alpha\},\{Y+\beta,B\}\}\\
&=(1+\CL_X)\de^B(Y+\beta)~.
\end{aligned}
\end{equation}
Thus, $\de^B(Y+\beta)$ transforms as an ordinary section of $TM\oplus T^*M$, namely with the action of diffeomorphisms. 

In Double Field Theory, we have a similar split of contributions as in~\eqref{eq:Dorfman_GG}:
\begin{equation}
 \hat \CL^0_XY=X^M\dpar_M Y - Y^M\dpar_M X+\theta^M Y^K\dpar_M X_K
\end{equation}
for $X,Y\in \CCC^\infty_1(\CCE_2)$. Here, the first term describes the action of infinitesimal diffeomorphisms on the generalized vectors $Y$ regarded as functions and the second and third term describe the terms of the form $\de^{\delta B} Y$ for
\begin{equation}
 \delta B=\tfrac12 \dpar_M X_N \theta^M\theta^N~.
\end{equation}
We thus interpret the generalized Lie derivative as a transformation
\begin{equation}
 Y~~\xmapsto{~~\hat \CL_X~~}~~Y'=Y+\CL_XY-\iota_Y \dd X~,
\end{equation}
which amounts to the diffeomorphism
\begin{equation}
 \de^B Y=Y~~\xmapsto{~~\hat\CL_X~~}~~\de^{B'} Y'=Y+X^M\dpar_MY~.
\end{equation}
This transformation is readily checked in the case $B=0$ and $B'=\delta B$, and it is the analogue of the above situation in Generalized Geometry.

Regarding $B$-transformed doubled vectors as scalars on the doubled space may be unusual, but our interpretation is supported by the following points. First, we saw that the action of the generalized Lie derivative on functions $f\in\CC^\infty_0(\CE_2)$ generates the diffeomorphisms group on the base space of $\CE_2$. Second, we also know that the action of the generalized Lie derivative on arbitrary functions on $\CE_2$ induces a subgroup of the symplectomorphisms on $\CE_2$. The only remaining candidate is thus the group of diffeomorphisms. Third, this fits with the strong section condition, $\dpar_M X \dpar^M Y=0$ on arbitrary tensors $X,Y$, which always seemed incompatible with covariance from a differential geometric point of view. Fourth, doubled vectors combine vector fields and 1-forms, which transform in opposite ways under diffeomorphisms. The scalar transformation law can be regarded as a consistent ``compromise.''

The situation for an arbitrary initial Kalb--Ramond field $B$ is more involved. First of all, $\de^B$ does not truncate,
\begin{equation}
\begin{aligned}
 \de^B Y&=Y+\{Y,B\}+\tfrac12 \{\{Y,B\},B\}+\dots~,\\
 &=\theta_M(Y^M+Y^NB_N{}^M +\tfrac12 Y^KB_K{}^N B_N{}^M+\dots)\\
 &=\theta_M Y^N(\delta^M_N+B_N{}^M +\tfrac12 B_N{}^K B_K{}^M+\dots)~,
\end{aligned}
\end{equation}
but this expression converges: it is simply the usual matrix exponential multiplying $Y$. The problem is rather that in Double Field Theory, the analogue of the transformation $\de^B$ in Generalized Geometry which moves the $B$-field modification from the transition function to the homological vector field (and thus into the generalized Lie derivative) may now produce noncommutative or nonassociative spaces, which cannot be described without generalizing our framework.

\subsection{Specialization to \texorpdfstring{$\CCE_K$}{EK}}

Let us briefly specify the above discussion to $\CCE_K$; the necessary adaptions are rather straightforward. The generalized Lie derivative reads as 
\begin{equation}
\begin{aligned}
 \hat \CL_X Y&=\{\{\CQ,X\},Y\}=[X,Y]^\tau+Y_A\dd^\tau X^A+\\
 &~~+j\big((Y^1X^2-X^1Y^2)\theta^4_++(Y^2X_4-X^2Y_4)\xi^1+(X^1Y_4-Y^1 X_4)\xi^2\big)\\
 &~~+k\big((Y^1X^2-X^1Y^2)\theta^3_++k(Y^2X_3-X^2Y_3)\xi^1+k(X^1Y_3-Y^1 X_3)\xi^2\big)~,
\end{aligned}
\end{equation}
where we used again the notation introduced in~\eqref{eq:index_notation} and $X,Y\in \CC^\infty_{{\rm loc},1}(\CCE_K)$. 

The relevant section conditions~\eqref{eq:new_section_1} and~\eqref{eq:new_section_2} translate to 
\begin{equation}
\begin{aligned}
\{\{\CQ,\{\CQ,f\}\},Y\}&=\{\{\tfrac12\{\CQ,\CQ\},f\},Y\}\\
&=\{\{2p_3p_4,f\},Y\}\\
&=(\dpar_3 f)(\dpar_4 Y)+(\dpar_4 Y)(\dpar_3 f)=(\dpar_\alpha f)(\dpar^\alpha Y)=0~,
\end{aligned}
\end{equation}
and
\begin{equation}
\begin{aligned}
 \{\{\{\{\CQ,\CQ\},X\},Y\},Z\}&=\{\{\{p_3p_4,X\},Y\},Z\}\\
 &=\{(\dpar_\alpha X)(\dpar^\alpha Y),Z\}+\{\{\dpar_\alpha X,Y\},\dpar^\alpha Z\}=0~.
\end{aligned}
\end{equation}
Both conditions are trivially satisfied for the $L_\infty$-structures $\sL^{\rm loc}_{\hat \pr_1}$ and $\sL^{\rm loc}_{\hat \pr_2}$ defined in~\eqref{eq:L_infty_algebra_structures}.

The generalized Lie derivative now acts as the semi-direct product of diffeomorphisms on the correspondence space $K$ and closed 2-forms along the directions $x^1$ and $x^2$ not involved in the T-duality.

\section{Discussion}

We would like to stress that nilmanifolds as base spaces for a discussion of global Double Field Theory were chosen merely for pedagogical reasons. If we have a T-duality relation between two spaces $X_1$ and $X_2$ fibered over a common base manifold $M$ and carrying $B$-fields or, equivalently, gerbes $\CCG_1$ and $\CCG_2$ with connective structure, we always have a correspondence space $K$ carrying a gerbe $\CCG_K$ which is the tensor product of the pullbacks of $\CCG_1$ and $\CCG_2$. This gerbe comes with a Courant algebroid $\CCC_K$, which can be restricted to a pre-N$Q$-manifold $\CCE_K$ capturing the global geometry of Double Field Theory:
\vspace*{0.5cm}
\begin{equation}
 \myxymatrix{
     & \CCG_K\ar@{->}[dr] \ar@{<.>}@/^4ex/[rr]& \CCE_K\ar@{->}[d] & \CCC_K \ar@{->}[dl] \ar@{->}[l] & & \\
     \CCC_{1}\ar@{<.>}[d] \ar@{->}[dr] & & K:=X_1\times_{M}X_2 \ar@{->}[dl]^{\pr_1} \ar@{->}[dr]_{\pr_2}& & \CCC_{2} \ar@{->}[dl]\ar@{<.>}[d]\\
     \CCG_{1} \ar@{->}[r] & X_1\ar@{->}[dr]_{\pi_1} & & X_2\ar@{->}[dl]^{\pi_2} & \CCG_2 \ar@{->}[l] \\
     & & M & &
    }
\end{equation}
In this way, any geometric T-duality\footnote{not involving noncommutative or nonassociative spaces} is readily captured in our framework as the special case in which the pull-backs of the Dixmier--Douady classes of the gerbes to $K$ agree.

The extension to Exceptional Field Theory (EFT), however, is more problematic. The issue here is that a clear analogue of the local picture developed in~\cite{Deser:2016qkw} is still missing. Note that the analogues of Courant algebroids capturing the symmetries of the $C$-field are mostly clear. They are higher versions of Vinogradov algebroids containing shifted copies to accommodate both M2- and M5-brane windings, cf.~\cite{Baraglia:2011dg,Arvanitakis:2018cyo}. The situation over the correspondence space, however, is much less clear. In particular, we do not have (yet) the higher analogue of $\CCE_K$, not even in the local case.

\section*{Acknowledgments}

This work has been partially supported by the European Cooperation in Science and Technology (COST) Action MP1405 QSPACE. The research of AD was supported by OP RDE project No.~CZ.02.2.69/0.0/0.0/16\_027/0008495, International Mobility of Researchers at Charles University.

\bibliography{bigone}

\begin{thebibliography}{10}

\bibitem{Aldazabal:2013sca}
G.~Aldazabal, D.~Marques, and C.~Nunez,
{\em Double field theory: A pedagogical review,}
\href{http://dx.doi.org/10.1088/0264-9381/30/16/163001}{Class. Quant. Grav.
  {\bf 30} (2013) 163001} [{\tt
  \href{http://www.arxiv.org/abs/1305.1907}{1305.1907 [hep-th]}}].

\bibitem{Berman:2013eva}
D.~S.~Berman and D.~C.~Thompson,
{\em {Duality symmetric string and M-theory},}
\href{http://dx.doi.org/10.1016/j.physrep.2014.11.007}{Phys. Rept. {\bf 566}
  (2014)~1} [{\tt \href{http://www.arxiv.org/abs/1306.2643}{1306.2643
  [hep-th]}}].

\bibitem{Hohm:2013bwa}
O.~Hohm, D.~L{\"u}st, and B.~Zwiebach,
{\em {The spacetime of double field theory: Review, remarks, and outlook},}
\href{http://dx.doi.org/10.1002/prop.201300024}{Fortsch. Phys. {\bf 61} (2013)
  926} [{\tt \href{http://www.arxiv.org/abs/1309.2977}{1309.2977 [hep-th]}}].

\bibitem{Cederwall:2014kxa}
M.~Cederwall,
{\em {The geometry behind double geometry},}
\href{http://dx.doi.org/10.1007/JHEP09(2014)070}{JHEP {\bf 1409} (2014) 070}
  [{\tt \href{http://www.arxiv.org/abs/1402.2513}{1402.2513 [hep-th]}}].

\bibitem{Cederwall:2014opa}
M.~Cederwall,
{\em {T-duality and non-geometric solutions from double geometry},}
\href{http://dx.doi.org/10.1002/prop.201400069}{Fortsch. Phys. {\bf 62} (2014)
  942} [{\tt \href{http://www.arxiv.org/abs/1409.4463}{1409.4463 [hep-th]}}].

\bibitem{Blumenhagen:2014gva}
R.~Blumenhagen, F.~Hassler, and D.~L{\"u}st,
{\em Double field theory on group manifolds,}
\href{http://dx.doi.org/10.1007/JHEP02(2015)001}{JHEP {\bf 1502} (2015) 001}
  [{\tt \href{http://www.arxiv.org/abs/1410.6374}{1410.6374 [hep-th]}}].

\bibitem{Blumenhagen:2015zma}
R.~Blumenhagen, P.~du~Bosque, F.~Hassler, and D.~L{\"u}st,
{\em {Generalized metric formulation of double field theory on group
  manifolds},}
\href{http://dx.doi.org/10.1007/JHEP08(2015)056}{JHEP {\bf 1508} (2015) 056}
  [{\tt \href{http://www.arxiv.org/abs/1502.02428}{1502.02428 [hep-th]}}].

\bibitem{Hassler:2016srl}
F.~Hassler,
{\em The topology of Double Field Theory,}
\href{http://dx.doi.org/10.1007/JHEP04(2018)128}{JHEP {\bf 1804} (2018) 128}
  [{\tt \href{http://www.arxiv.org/abs/1611.07978}{1611.07978 [hep-th]}}].

\bibitem{Freidel:2017yuv}
L.~Freidel, F.~J.~Rudolph, and D.~Svoboda,
{\em Generalised kinematics for double field theory,}
\href{http://dx.doi.org/10.1007/JHEP11(2017)175}{JHEP {\bf 1711} (2017) 175}
  [{\tt \href{http://www.arxiv.org/abs/1706.07089}{1706.07089 [hep-th]}}].

\bibitem{Deser:2016qkw}
A.~Deser and C.~Saemann,
{\em {Extended Riemannian geometry I: Local double field theory},}
\href{http://dx.doi.org/10.1007/s00023-018-0694-2}{Ann. H. Poincare. {\bf 19}
  (2018) 2297} [{\tt \href{http://www.arxiv.org/abs/1611.02772}{1611.02772
  [hep-th]}}].

\bibitem{Deser:2017fko}
A.~Deser, M.~A.~Heller, and C.~Saemann,
{\em {Extended Riemannian geometry II: Local heterotic double field theory},}
\href{http://dx.doi.org/10.1007/JHEP04(2018)106}{JHEP {\bf 1804} (2018) 106}
  [{\tt \href{http://www.arxiv.org/abs/1711.03308}{1711.03308 [hep-th]}}].

\bibitem{Gualtieri:2003dx}
M.~Gualtieri,
{\em {Generalized complex geometry},} PhD thesis, Oxford (2003)
[{\tt \href{http://www.arxiv.org/abs/math.DG/0401221}{math.DG/0401221}}].

\bibitem{Roytenberg:0203110}
D.~Roytenberg,
{\em On the structure of graded symplectic supermanifolds and Courant
  algebroids,}
in: ``Quantization, Poisson Brackets and Beyond,'' ed.\ Theodore Voronov,
  Contemp. Math., Vol. 315, Amer. Math. Soc., Providence, RI, 2002
[{\tt \href{http://www.arxiv.org/abs/math.SG/0203110}{math.SG/0203110}}].

\bibitem{Deser:2014mxa}
A.~Deser and J.~Stasheff,
{\em {Even symplectic supermanifolds and double field theory},}
\href{http://dx.doi.org/10.1007/s00220-015-2443-4}{Commun. Math. Phys. {\bf
  339} (2015) 1003} [{\tt \href{http://www.arxiv.org/abs/1406.3601}{1406.3601
  [math-ph]}}].

\bibitem{Deser:2018oyg}
A.~Deser and C.~Saemann,
{\em {Derived brackets and symmetries in Generalized Geometry and Double Field
  Theory},}
\href{http://inspirehep.net/record/1658630/files/1803.01659.pdf}{PoS {\bf
  CORFU2017} (2018) 141} [{\tt
  \href{http://www.arxiv.org/abs/1803.01659}{1803.01659 [hep-th]}}].

\bibitem{Fiorenza:0601312}
D.~Fiorenza and M.~Manetti,
{\em $L_\infty$ structures on mapping cones,}
\href{http://dx.doi.org/10.2140/ant.2007.1.301}{Alg. Numb. Th. {\bf 1} (2007)
  301} [{\tt
  \href{http://www.arxiv.org/abs/math.QA/0601312}{math.QA/0601312}}].

\bibitem{Getzler:1010.5859}
E.~Getzler,
{\em Higher derived brackets,}
{\tt \href{http://www.arxiv.org/abs/1010.5859}{1010.5859 [math-ph]}}.

\bibitem{Berman:2014jba}
D.~S.~Berman, M.~Cederwall, and M.~J.~Perry,
{\em {Global aspects of double geometry},}
\href{http://dx.doi.org/10.1007/JHEP09(2014)066}{JHEP {\bf 1409} (2014) 066}
  [{\tt \href{http://www.arxiv.org/abs/1401.1311}{1401.1311 [hep-th]}}].

\bibitem{Hohm:2012gk}
O.~Hohm and B.~Zwiebach,
{\em {Large gauge transformations in double field theory},}
\href{http://dx.doi.org/10.1007/JHEP02(2013)075}{JHEP {\bf 1302} (2013) 075}
  [{\tt \href{http://www.arxiv.org/abs/1207.4198}{1207.4198 [hep-th]}}].

\bibitem{Hull:2014mxa}
C.~M.~Hull,
{\em {Finite gauge transformations and geometry in double field theory},}
\href{http://dx.doi.org/10.1007/JHEP04(2015)109}{JHEP {\bf 1504} (2015) 109}
  [{\tt \href{http://www.arxiv.org/abs/1406.7794}{1406.7794 [hep-th]}}].

\bibitem{Park:2013mpa}
J.-H.~Park,
{\em Comments on double field theory and diffeomorphisms,}
\href{http://dx.doi.org/10.1007/JHEP06(2013)098}{JHEP {\bf 1306} (2013) 098}
  [{\tt \href{http://www.arxiv.org/abs/1304.5946}{1304.5946 [hep-th]}}].

\bibitem{Papadopoulos:2014mxa}
G.~Papadopoulos,
{\em {Seeking the balance: Patching double and exceptional field theories},}
\href{http://dx.doi.org/10.1007/JHEP10(2014)089}{JHEP {\bf 1410} (2014) 089}
  [{\tt \href{http://www.arxiv.org/abs/1402.2586}{1402.2586 [hep-th]}}].

\bibitem{Heller:2016abk}
M.~A.~Heller, N.~Ikeda, and S.~Watamura,
{\em {Unified picture of non-geometric fluxes and T-duality in double field
  theory via graded symplectic manifolds},}
\href{http://dx.doi.org/10.1007/JHEP02(2017)078}{JHEP {\bf 1702} (2017) 078}
  [{\tt \href{http://www.arxiv.org/abs/1611.08346}{1611.08346 [hep-th]}}].

\bibitem{Bouwknegt:2003vb}
P.~Bouwknegt, J.~Evslin, and V.~Mathai,
{\em T-Duality: Topology change from H-flux,}
\href{http://dx.doi.org/10.1007/s00220-004-1115-6}{Commun. Math. Phys. {\bf
  249} (2004) 383} [{\tt
  \href{http://www.arxiv.org/abs/hep-th/0306062}{hep-th/0306062}}].

\bibitem{Cavalcanti:2011wu}
G.~R.~Cavalcanti and M.~Gualtieri,
{\em Generalized complex geometry and T-duality,}
in: ``A Celebration of the Mathematical Legacy of Raoul Bott,'' CRM Proceedings
  \& Lecture Notes, American Mathematical Society, 2010
[{\tt \href{http://www.arxiv.org/abs/1106.1747}{1106.1747 [math.DG]}}].

\bibitem{Hohm:2017pnh}
O.~Hohm and B.~Zwiebach,
{\em {$L_{\infty}$ algebras and field theory},}
\href{http://dx.doi.org/10.1002/prop.201700014}{Fortsch. Phys. {\bf 65} (2017)
  1700014} [{\tt \href{http://www.arxiv.org/abs/1701.08824}{1701.08824
  [hep-th]}}].

\bibitem{Jurco:2018sby}
B.~Jurco, L.~Raspollini, C.~Saemann, and M.~Wolf,
{\em {$L_\infty$-algebras of classical field theories and the
  Batalin--Vilkovisky formalism},}
{\tt \href{http://www.arxiv.org/abs/1809.09899}{1809.09899 [hep-th]}}.

\bibitem{Hull:2009sg}
C.~M.~Hull and R.~A.~Reid{--}Edwards,
{\em {Non-geometric backgrounds, doubled geometry and generalised T-duality},}
\href{http://dx.doi.org/10.1088/1126-6708/2009/09/014}{JHEP {\bf 0909} (2009)
  014} [{\tt \href{http://www.arxiv.org/abs/0902.4032}{0902.4032 [hep-th]}}].

\bibitem{Hitchin:1999fh}
N.~J.~Hitchin,
{\em {Lectures on special Lagrangian submanifolds},}
{\tt \href{http://www.arxiv.org/abs/math.DG/9907034}{math.DG/9907034}}.

\bibitem{Bunk:2016rta}
S.~Bunk, C.~Saemann, and R.~J.~Szabo,
{\em {The 2-Hilbert space of a prequantum bundle gerbe},}
\href{http://dx.doi.org/10.1142/S0129055X18500010}{Rev. Math. Phys. {\bf 30}
  (2018) 1850001} [{\tt \href{http://www.arxiv.org/abs/1608.08455}{1608.08455
  [math-ph]}}].

\bibitem{Bressler:2002ur}
P.~Bressler and A.~Chervov,
{\em {Courant algebroids},}
\href{http://dx.doi.org/10.1007/s10958-005-0251-7}{J. Math. Sci. {\bf 128}
  (2005) 3030} [{\tt
  \href{http://www.arxiv.org/abs/hep-th/0212195}{hep-th/0212195}}].

\bibitem{Collier:1108.1525}
B.~L.~Collier,
{\em Infinitesimal symmetries of Dixmier--Douady gerbes,}
{\tt \href{http://www.arxiv.org/abs/1108.1525}{1108.1525 [math.DG]}}.

\bibitem{Hitchin:2005in}
N.~Hitchin,
{\em {Brackets, forms and invariant functionals},}
{\tt \href{http://www.arxiv.org/abs/math.DG/0508618}{math.DG/0508618}}.

\bibitem{Chatzistavrakidis:2013wra}
A.~Chatzistavrakidis, L.~Jonke, and O.~Lechtenfeld,
{\em Dirac structures on nilmanifolds and coexistence of fluxes,}
\href{http://dx.doi.org/10.1016/j.nuclphysb.2014.03.013}{Nucl. Phys. B {\bf
  883} (2014)~59} [{\tt \href{http://www.arxiv.org/abs/1311.4878}{1311.4878
  [hep-th]}}].

\bibitem{Baraglia:2011dg}
D.~Baraglia,
{\em {Leibniz algebroids, twistings and exceptional generalized geometry},}
\href{http://dx.doi.org/10.1016/j.geomphys.2012.01.007}{J. Geom. Phys. {\bf 62}
  (2012) 903} [{\tt \href{http://www.arxiv.org/abs/1101.0856}{1101.0856
  [math.DG]}}].

\bibitem{Arvanitakis:2018cyo}
A.~S.~Arvanitakis,
{\em {Brane Wess--Zumino terms from AKSZ and exceptional generalised geometry
  as an $L_\infty$-algebroid},}
{\tt \href{http://www.arxiv.org/abs/1804.07303}{1804.07303 [hep-th]}}.

\end{thebibliography}

\bibliographystyle{latexeu}

\end{document}